\documentclass[
reprint,
superscriptaddress,
showpacs,preprintnumbers,
bibnotes,
amsmath,amssymb,
aps,
prl,
floatfix,
]{revtex4-2}

\usepackage[colorlinks=true,citecolor=magenta]{hyperref}
\usepackage{orcidlink}
\usepackage{graphicx}
\usepackage{color}
\usepackage{comment}

\begin{document}

\title{Testing strong-field QED to second-order in the highly correlated atomic system berylliumlike Pb$^{78+}$ by electron-ion collision spectroscopy}


\author{S.~Schippers\,\orcidlink{0000-0002-6166-7138}} 
\affiliation{I.~Physikalisches Institut, Justus-Liebig-Universit\"{a}t Gie{\ss}en, 35392 Giessen, Germany}
\affiliation{Helmholtz Forschungsakademie Hessen f\"ur FAIR (HFHF), GSI Helmholtzzentrum f\"ur Schwerionenforschung, 64291 Darmstadt, Germany}

\author{C.~Brandau\,\orcidlink{0000-0001-8825-8820}}
\affiliation{I.~Physikalisches Institut, Justus-Liebig-Universit\"{a}t Gie{\ss}en, 35392 Giessen, Germany}
\affiliation{GSI Helmholtzzentrum f\"ur Schwerionenforschung GmbH, 64291 Darmstadt, Germany}

\author{S.~Fuchs\,\orcidlink{0009-0004-3560-1049}}
\affiliation{I.~Physikalisches Institut, Justus-Liebig-Universit\"{a}t Gie{\ss}en, 35392 Giessen, Germany}
\affiliation{Helmholtz Forschungsakademie Hessen f\"ur FAIR (HFHF), GSI Helmholtzzentrum f\"ur Schwerionenforschung, 64291 Darmstadt, Germany}

\author{M.~Lestinsky\,\orcidlink{0000-0001-7384-1917}}
\affiliation{GSI Helmholtzzentrum f\"ur Schwerionenforschung GmbH, 64291 Darmstadt, Germany}

\author{S.~X.~Wang\,\orcidlink{0000-0002-6305-3762}}
\affiliation{I.~Physikalisches Institut, Justus-Liebig-Universit\"{a}t Gie{\ss}en, 35392 Giessen, Germany}
\affiliation{Helmholtz Forschungsakademie Hessen f\"ur FAIR (HFHF), GSI Helmholtzzentrum f\"ur Schwerionenforschung, 64291 Darmstadt, Germany}

\author{C.~Y.~Zhang\,\orcidlink{0000-0003-1935-6907}}
\affiliation{Department of Physics, University of Strathclyde, Glasgow G4 0NG, UK}

\author{N.~R.~Badnell\,\orcidlink{0000-0001-7418-7996}}
\affiliation{Department of Physics, University of Strathclyde, Glasgow G4 0NG, UK}

\author{A.~Borovik~Jr.\,\orcidlink{0000-0002-6952-3550}}
\affiliation{I.~Physikalisches Institut, Justus-Liebig-Universit\"{a}t Gie{\ss}en, 35392 Giessen, Germany}
\affiliation{Helmholtz-Institut Jena,  07743 Jena, Germany}

\author{M.~Fogle\,\orcidlink{0000-0003-3511-262X}}
\affiliation{Department of Physics, Auburn University, Alabama 36849, USA}

\author{V.~Hannen\,\orcidlink{0000-0002-2944-8373}}
\affiliation{Institut f\"ur Kernphysik, Universit\"at M\"unster, 48149 M\"unster, Germany}

\author{Z.~Harman}
\affiliation{Max-Planck-Institut f\"ur Kernphysik, 69117 Heidelberg, Germany}

\author{P.-M.~Hillenbrand\,\orcidlink{0000-0003-0166-2666}}
\affiliation{I.~Physikalisches Institut, Justus-Liebig-Universit\"{a}t Gie{\ss}en, 35392 Giessen, Germany}
\affiliation{GSI Helmholtzzentrum f\"ur Schwerionenforschung GmbH, 64291 Darmstadt, Germany}

\author{E.~B.~Menz\,\orcidlink{0000-0003-1451-7089}}
\affiliation{GSI Helmholtzzentrum f\"ur Schwerionenforschung GmbH, 64291 Darmstadt, Germany}
\affiliation{Helmholtz-Institut Jena,  07743 Jena, Germany}

\author{Y. Zhang}
\affiliation{Max-Planck-Institut f\"ur Kernphysik, 69117 Heidelberg, Germany}
\affiliation{School of Science, Xi'an Jiaotong University, Xi'an 710049, China}

\author{Z.~Andelkovic\,\orcidlink{0009-0003-0576-4317}}
\affiliation{GSI Helmholtzzentrum f\"ur Schwerionenforschung GmbH, 64291 Darmstadt, Germany}

\author{F.~Herfurth}
\affiliation{GSI Helmholtzzentrum f\"ur Schwerionenforschung GmbH, 64291 Darmstadt, Germany}

\author{R.~He{\ss}\,\orcidlink{0000-0003-1239-2585}}
\affiliation{GSI Helmholtzzentrum f\"ur Schwerionenforschung GmbH, 64291 Darmstadt, Germany}

\author{A.~Kalinin\,\orcidlink{0000-0001-7710-1799}}
\affiliation{GSI Helmholtzzentrum f\"ur Schwerionenforschung GmbH, 64291 Darmstadt, Germany}

\author{C.~Kozhuharov\,\orcidlink{0000-0001-9487-5681}}
\affiliation{GSI Helmholtzzentrum f\"ur Schwerionenforschung GmbH, 64291 Darmstadt, Germany}

\author{C.~Krantz\,\orcidlink{0000-0003-1727-8319}}
\affiliation{GSI Helmholtzzentrum f\"ur Schwerionenforschung GmbH, 64291 Darmstadt, Germany}

\author{S.~Litvinov\,\orcidlink{0000-0002-7055-6763}}
\affiliation{GSI Helmholtzzentrum f\"ur Schwerionenforschung GmbH, 64291 Darmstadt, Germany}

\author{B.~Lorentz\,\orcidlink{0000-0002-1086-3737}}
\affiliation{GSI Helmholtzzentrum f\"ur Schwerionenforschung GmbH, 64291 Darmstadt, Germany}

\author{U.~Spillmann\,\orcidlink{0000-0001-7281-5063}}
\affiliation{GSI Helmholtzzentrum f\"ur Schwerionenforschung GmbH, 64291 Darmstadt, Germany}

\author{M.~Steck}
\affiliation{GSI Helmholtzzentrum f\"ur Schwerionenforschung GmbH, 64291 Darmstadt, Germany}

\author{G.~Vorobyev}
\affiliation{GSI Helmholtzzentrum f\"ur Schwerionenforschung GmbH, 64291 Darmstadt, Germany}

\author{D.~Bana\'s\,\orcidlink{0000-0003-1566-5446}}
\affiliation{Institute of Physics, Jan Kochanowski University, 25-406 Kielce, Poland}

\author{S.~Fritzsche\,\orcidlink{0000-0003-3101-2824}}
\affiliation{GSI Helmholtzzentrum f\"ur Schwerionenforschung GmbH, 64291 Darmstadt, Germany}
\affiliation{Helmholtz-Institut Jena,  07743 Jena, Germany}
\affiliation{Theoretisch-Physikalisches Institut, Friedrich-Schiller-Universit\"at Jena, 07743 Jena, Germany}

\author{E.~Lindroth\,\orcidlink{0000-0003-3444-1317}}
\affiliation{Department of Physics, Stockholm University, AlbaNova, 10691 Stockholm, Sweden}

\author{X.~Ma\,\orcidlink{0000-0001-9831-0565}}
\affiliation{Institute of Modern Physics, Chinese Academy of Sciences, 730000 Lanzhou, China}

\author{A.~M\"uller\,\orcidlink{0000-0002-0030-6929}}
\affiliation{I.~Physikalisches Institut, Justus-Liebig-Universit\"{a}t Gie{\ss}en, 35392 Giessen, Germany}

\author{R.~Schuch\,\orcidlink{0000-0002-5843-0586}}
\affiliation{Department of Physics, Stockholm University, AlbaNova, 10691 Stockholm, Sweden}

\author{A.~Surzhykov\,\orcidlink{0000-0002-6441-0864}}
\affiliation{Physikalisch–Technische Bundesanstalt, 38116 Braunschweig, Germany}
\affiliation{Institut f\"ur Mathematische Physik, Technische Universität Braunschweig, 38106 Braunschweig, Germany}

\author{M.~Trassinelli\,\orcidlink{0000-0003-4414-1801}}
\affiliation{Institut des NanoSciences de Paris, CNRS, Sorbonne Universit\'e, 75005 Paris, France}

\author{K.~Ueberholz}
\affiliation{Institut f\"ur Kernphysik, Universit\"at M\"unster, 48149 M\"unster, Germany}

\author{C.~Weinheimer\,\orcidlink{0000-0002-4083-9068}}
\affiliation{Institut f\"ur Kernphysik, Universit\"at M\"unster, 48149 M\"unster, Germany}

\author{Th.~Stöhlker\,\orcidlink{0000-0003-0461-3560}}
\affiliation{GSI Helmholtzzentrum f\"ur Schwerionenforschung GmbH, 64291 Darmstadt, Germany}
\affiliation{Helmholtz-Institut Jena, 07743 Jena, Germany}
\affiliation{Institut f\"ur Optik und Quantenelektronik, Friedrich-Schiller-Universit\"at Jena, 07743 Jena, Germany}


\date{\today}

\begin{abstract}
A low-energy storage ring with an ultracold electron cooler has been coupled with a heavy-ion accelerator facilitating high-resolution electron-ion collision spectroscopy of the heaviest few-electron ions. In the present work resonant electron-ion recombination of berylliumlike Pb$^{78+}$ ions was measured in the collision-energy range 9.3--16.5~eV and a value of 244.937(30)~eV is derived for the Pb$^{78+}$($2s^2\;^1S_0 - 2s\,2p\;^3P_1$) excitation energy. This result agrees with the most recent (less accurate) theoretical value of 244.942(52)~eV [Malyshev et~al., Physical Review A 110, 062824 (2024)], which has been calculated by applying strong-field QED rigorously up to the second order. The present investigation suggests that further technical improvements can potentially increase the experimental accuracy by an order of magnitude.
\end{abstract}

\maketitle

The experimental technique of electron-ion collision spectroscopy of dielectronic recombination (DR) resonances  is a very successful approach for studying the properties of highly charged ions.  The range of topics that has been addressed by collision spectroscopy comprises QED tests in strong-fields \cite{Lindroth2001,Brandau2003b,Lestinsky2008a}, investigations of nuclear properties via isotope shift or hyperfine splitting \cite{Schuch2005a,Lestinsky2008a,Brandau2008a}, and lifetime studies \cite{Schmidt1994,Schippers2007a}. Recently, a major technological leap was achieved by moving the low-energy heavy-ion storage ring CRYRING \cite{Abrahamsson1993a} with its ultra-cold electron cooler \cite{Danared1994} from its original location at Stockholm, Sweden, to the international Facility for Antiprotron and Ion Research (FAIR) on the premises of the GSI Helmholtzzentrum f\"ur Schwerionenforschung in Darmstadt, Germany.  This world-unique arrangement facilitates the storage heavy highly charged ions at low ion energies and their cooling with a very cold electron beam such that spectroscopic techniques can be applied with highest precision \cite{Madzunkov2002,Fogle2003a,Lestinsky2016,Zhu2022,Brandau2025}.

The present work focuses on the collision spectroscopy of a very heavy berylliumlike ion, $^{208}$Pb$^{78+}$. In particular, we have accurately measured the positions of $2s\,2p\,(^3P_1)\,19l_j$ DR resonances which are associated with the $2s^2\;^1S_0 - 2s\,2p\;^3P_1$ excitation of the berylliumlike core and the capture of the initially free electron into the $n=19$ shell. As we will demonstrate, the binding energy $E_b(nl_j)$ of the $n=19$ Rydberg electron can be calculated to very high precision, such that the $2s^2\;^1S_0 - 2s\,2p\;^3P_1$ excitation energy $E_\mathrm{exci}$ can be obtained from the measured DR resonance energies
\begin{equation}\label{eq:Eres}
    E_\mathrm{DR} = E_\mathrm{exci}-E_b(nl_j)
\end{equation}
with an uncertainty of only 30~meV in the present case.

From a fundamental point of view, Be-like ions are of particular interest for atomic physics because correlation effects play a relatively strong role in  these four-electron systems. On the one hand, this greatly complicates ab-initio QED calculations beyond the first-order approximation \cite{Malyshev2021}. On the other hand, Be-like ions still have a comparatively simple electronic structure and, thus, provide ideal test cases to benchmark theory.

Experimental results for Be-like ions heavier than xenon ($Z\geq54$) are scarce because very heavy few-electron ions are not easily accessible.  The U$^{88+}$($2s^2\;^1S_0\; - 2s\,2p\;^1P_1$) and U$^{88+}$($2s^2\;^1S_0\; - 2s\,2p\;^3P_1$) excitation energies have been measured using x-ray spectroscopy  at the Livermore SuperEBIT \cite{Beiersdorfer1993,Beiersdorfer2005a}  with uncertainties of about 210~meV and 12~meV, respectively. The U$^{88+}$($2s^2\;^1S_0\; - 2s\,2p\;^1P_1$) transition energy was also measured at the Darmstadt high-energy storage ring ESR \cite{Loetzsch2024} with an uncertainty of 160~meV.  For xenon, the $2s^2\;^1S_0\; - 2s\,2p\;^3P_1$ intercombination line was studied by using  beam-foil spectroscopy \cite{Moeller1989,Buettner1992,Feili2005} and by observing x-ray lines from Xe$^{50+}$ ions confined in an electron beam ion trap (EBIT) \cite{Traebert2003}. More recently, energies for several fine-structure components of the $2s^2 - 2s\,2p$ transition in Xe$^{50+}$ were derived with high accuracy from a collision-spectroscopy experiment at the ESR \cite{Bernhardt2015a}. 

In the past four decades, calculations of energy levels of Be-like ions have been performed with different theoretical approaches \cite[e.g.,][]{Cheng1979,Safronova1996b,Gu2005a,Cheng2008a}.  More recent studies have taken higher-order QED effects, nuclear mass, nuclear size, and nuclear polarization effects, as well as the frequency-dependent Breit interaction into account \cite{Malyshev2014,Malyshev2015,Kaygorodov2019}. The latest theoretical efforts include rigorously all second-order QED contributions, i.e., two-photon exchange, two-loop and QED screening diagrams \cite{Malyshev2021,Malyshev2023,Malyshev2024}. For Be-like Xe$^{50+}$, the resulting excitation energies deviate from the beam-foil spectroscopy and EBIT values \cite{Feili2005,Traebert2003}, but agree within the experimental uncertainties with our earlier results from electron-ion collision spectroscopy at the ESR \cite{Bernhardt2015a}. As explained in much detail below, the corresponding value from  the present study with much more highly charged Pb$^{78+}$ is 30~meV, i.e., even lower than the 52-meV uncertainty quoted for the most recent theoretical value of the  Pb$^{78+}$($2s^2\;^1S_0 - 2s\,2p\;^3P_1$) transition energy \cite{Malyshev2024}. The present work, thus, represents a stringent experimental test of the  presently still untested high-order QED contributions for a highly correlated atomic system.

The present experiment was the first CRYRING experiment that made use of the full GSI accelerator chain --- comprising the linear accelerator UNILAC, the heavy-ion synchrotron SIS-18, and the high-energy storage ring ESR --- for producing and injecting ions into the CRYRING, where the electron-ion spectroscopy measurements were performed (see \cite{Brandau2025} for a general description of the experimental setup and procedures). In SIS-18, on the average $10^9$ lead ions per pulse were accelerated to an energy of 54~MeV/nucleon. The desired charge state $q=78$ was obtained  by passing the ions through a stripper foil located in the extraction beam line of SIS-18, which was used to transfer the ions to the ESR. There, the  ions were cooled and decelerated to the final ion energy of 11.3~MeV/nucleon by operating ESR in synchrotron mode. Eventually, up to $6\cdot10^6$ $^{208}$Pb$^{78+}$ ions were injected into CRYRING, where they were stored and further cooled to reduce size and momentum spread of the ion beam  before the start of a measurement.

In the CRYRING electron cooler \cite{Danared1994,Krantz2021}, the electrons are magnetically guided. In the electron-ion interaction section the electrons move coaxially with the ions in the same direction. The diameters of the electron and ion beams were 23~mm and about 2~mm, respectively. With the electron and ion  velocities relative to the vacuum speed of light, $c$, being  $\beta_e=v_e/c$ and $\beta_i=v_i/c$, respectively, the electron-ion collision energy in the center-of-mass frame can be expressed as \cite{Wang2024}
\begin{equation}
	E_{\mathrm{cm}} = m_ic^{2} (1 + \mu) \left[ \sqrt{1 + \frac{2\mu(\gamma_\mathrm{rel} - 1)}{(1 + \mu)^{2}} } - 1 \right],\label{eq:Ecm}
\end{equation}
where  $\mu=m_e/m_i\ll1$ is the ratio of the electron rest mass $m_e$  and the ion rest mass $m_i$, and
\begin{equation}\label{eq:gammacm}
\gamma_\mathrm{rel} = \gamma_{e}\gamma_{i}(1 - \beta_e\beta_i\cos\theta)
\end{equation}
with $\gamma_e=(1-\beta_e^2)^{-1/2}$, $\gamma_i=(1-\beta_i^2)^{-1/2}$, and  the angle $\theta$  between the two interacting beams. 

At the electron-cooling condition, i.e., for $\theta=0$ and $\beta_i=\beta_e$, Eqs.~\eqref{eq:gammacm} and \eqref{eq:Ecm} yield $\gamma_\mathrm{rel} = 1$ and $E_\mathrm{cm}=0$~eV, respectively, as expected. The associated laboratory-frame electron energy is the cooling energy $E_\mathrm{cool}$, which is related to $\beta_i$ via $\gamma_i = (1-\beta_i^2)^{-1/2} = 1 + E_\mathrm{cool}/(m_ec^2)$. When the electron energy is detuned from the cooling energy by an amount $E_d$ this results in $\gamma_e = (1-\beta_e^2)^{-1/2} = 1 + (E_\mathrm{cool}+E_d)/(m_ec^2)$. 
In the experiment, the cooling energy was adjusted by choosing the cooler cathode voltage $U_c$ accordingly, and the detuning energy was realized by setting a fast high-voltage amplifier, which is connected in series with the cathode voltage supply \cite[][Fig.~1]{Brandau2025}, to the appropriate detuning voltage $U_d$. Only $U_d$ was varied during electron-energy scans.

Merged-beams recombination rate coefficients $\alpha_\mathrm{mb}\left(E_\mathrm{cm}(E_d,E_\mathrm{cool},\theta)\right)$ were measured with a $\pm14\%$ uncertainty by recording the number of recombined ions as a function of $E_d$ ($E_\mathrm{cool}$ and $\theta=0$ were kept fixed) and by appropriately normalizing the background-subtracted count rate on electron density ($\sim10^7$~cm$^{-3}$), number of stored ions ($\sim5\cdot10^6$), and geometric beam overlap (for details see \cite{Brandau2025,Wang2024}). In order to preserve the ion-beam quality, $E_d$ was set to zero in between two measurement steps with $E_d\neq0$. The duration of each cooling and each measurement step was 10~ms, and one energy scan comprised up to 1000 cooling and 1000 measurement steps such that the 20-s duration of the scan was shorter than the 30-s storage lifetime of the ion beam.  The scan was started 2~s after an injection of an ion pulse into the CRYRING allowing for the cooling of the ion-beam. Each predefined injection-cooling-and-scanning sequence was repeated for a number of times until acceptable counting statistics were reached.

\begin{figure}
	\centering{\includegraphics[width=\columnwidth]{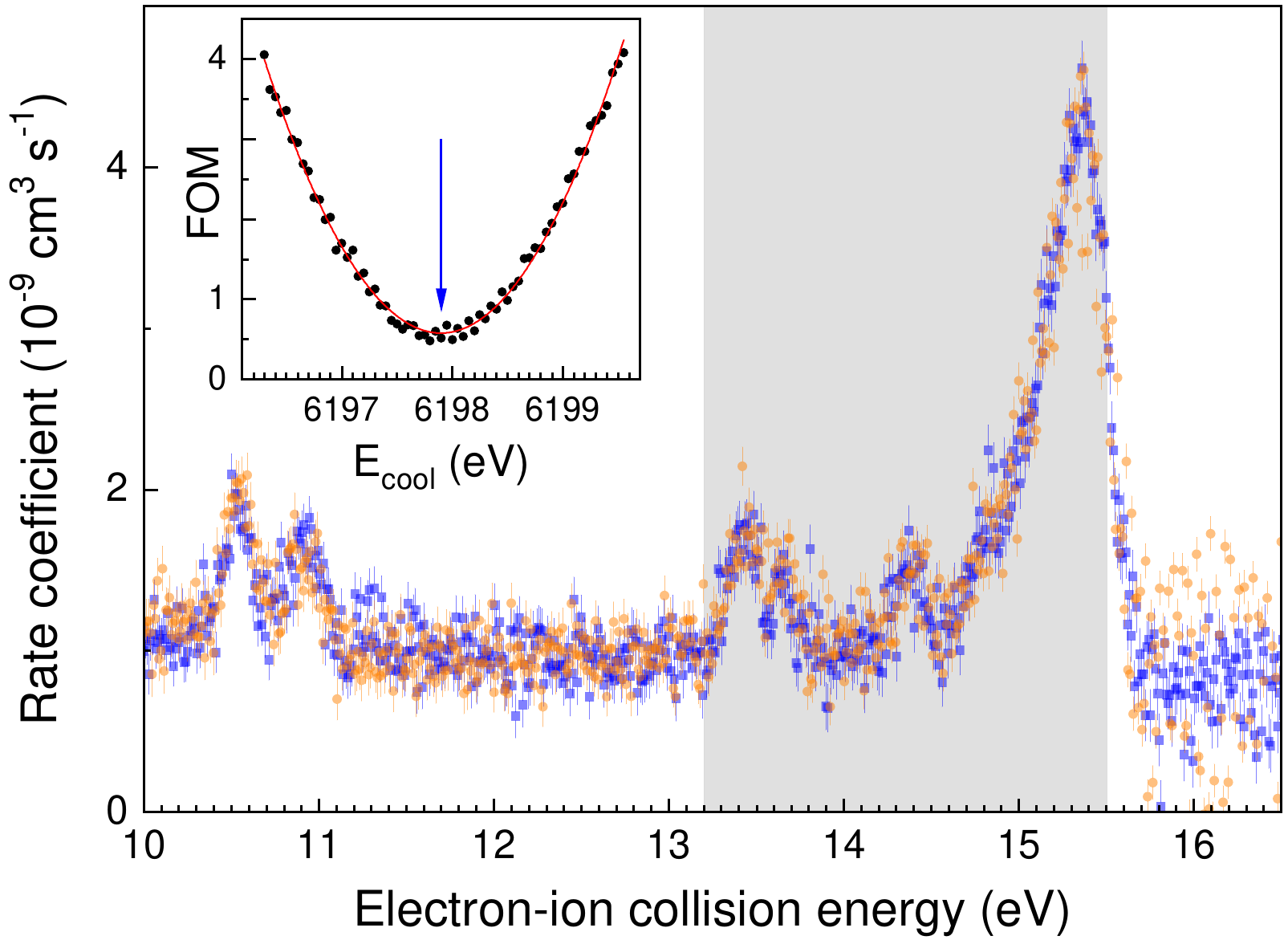}}
	\caption{\label{fig:slowfast}Two DR spectra measured with the electrons having been slower (blue squares) and faster (orange circles) than the ions. The gray shading marks the energy interval that was used for computing the FOM (see text).  The inset shows the variation of the FOM as a function of $E_\mathrm{cool}$. The optimum value is at $E_\mathrm{cool}=6197.90$~eV (marked by the vertical arrow) as determined by a fit of a parabola (full line) to the data points.}
\end{figure}

The cooling energy, $E_\mathrm{cool}$, was determined by exploiting a unique feature of the merged-beams arrangement, i.e., that the same electron-ion collision energy can be realized in two ways, first with the electrons being slower and second with the electrons being faster than the ions \cite{Fogle2003a}. These different conditions resulted in the two DR spectra that are shown in Fig.~\ref{fig:slowfast}. For the transformation of the laboratory electron energies to the center-of-mass frame (Eq.~\eqref{eq:Ecm}), $E_\mathrm{cool}$ has to be chosen such that identical resonance features appear at the same $E_\mathrm{cm}$.  For finding the correct $E_\mathrm{cool}$ value for the present experiment, the mean of the squared point-by-point differences was used as a figure of merit (FOM) for the agreement. As can be seen from the inset of Fig.~\ref{fig:slowfast}, it reaches its optimum at $E_\mathrm{cool}=6197.90$~eV. The same plot suggests that a conservative estimate for the associated uncertainty is $\Delta E_\mathrm{cool}=\pm0.2$~eV.

\begin{figure}
	\centering{\includegraphics[width=\columnwidth]{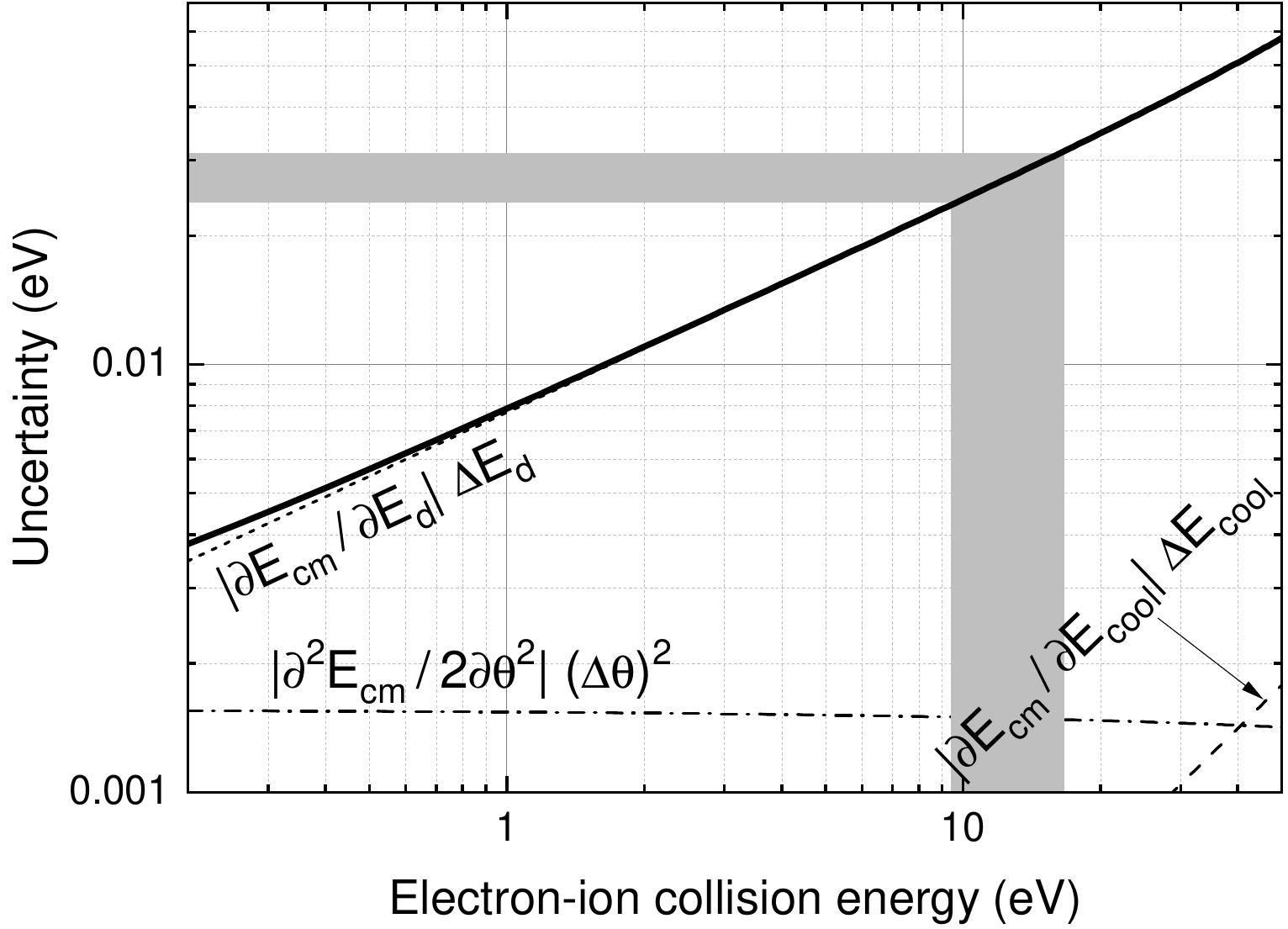}}
	\caption{\label{fig:DEcm}Total uncertainty $\Delta E_\mathrm{cm}$ (full line) of the electron-ion collision energy $E_\mathrm{cm}$ (Eq.~\eqref{eq:Ecm}) and the individual contributions by $\Delta E_\mathrm{cool}=\pm0.2$~eV (long-dashed line), $\Delta E_d=\{[10^{-4}e(U_c+U_d)]^2+[0.1e\,\delta U_\mathrm{sc}]^2\}^{1/2}$ (short-dashed line), and $\Delta\theta=\pm0.5$~mrad (dash-dotted line).  Since $\partial E_\mathrm{cm}/\partial\theta \propto\sin\theta$ vanishes for $\theta=0$ the second-order term of the Taylor expansion is used for accounting for $\Delta \theta$. The gray shaded area marks the energy range of the present experiment and the corresponding $\Delta E_\mathrm{cm}$ range of $24-31$~meV.}
\end{figure}

The  uncertainty of the detuning energy $E_d = -e[U_d - \delta U_\mathrm{sc}]$ depends on the uncertainties of the detuning voltage $U_d$ and of the electron beam's space charge potential $\delta U_\mathrm{sc}$. The latter is of minor importance. It is discussed in the appendix. In the experiment $U_c+U_d$ is measured online with a home-built high-voltage divider capable of following voltage changes on a sub-millisecond time scale \cite{Brandau2025}. This device has been calibrated off\-line against a more precise but slower high-voltage divider, which is a copy of the divider described in Ref.~\cite{Thuemmler2009} and which has a relative systematic uncertainty of maximally $8.4\cdot10^{-6}$ \cite{Rest2020}. This value includes the uncertainty of the 8.5-digits-voltmeter (Keysight 3458A DVM) that was used for readout. It is an order of magnitude lower than what must be attributed to the long-term stability of the fast HV divider, for which repeated calibration measurements yielded a  relative uncertainty of $10^{-4}$. Accounting also for the 10\% uncertainty of $\delta U_\mathrm{sc}= -1.7$~eV (see appendix) we arrive at $\Delta E_d=\{[10^{-4}e(U_c+U_d)]^2+[0.1e\,\delta U_\mathrm{sc}]^2\}^{1/2}$.

The uncertainty of the angle $\theta$ in Eq.~\eqref{eq:Ecm} amounts to $\Delta\theta = 0.5$~mrad. This was inferred from the beam-adjustment procedure, where the electron beam is magnetically steered such that the ion beam moves in the minimum of the electron beam's space-charge trough \cite{Boehm2001b}. When the electron beam is tilted from this configuration, the ion-beam is subject to a different mean electron space-charge potential, which changes its velocity. A variation of the angle by $\pm0.5$~mrad did not produce a visible change of the ions' frequency distribution as monitored by the Schottky ion-beam analysis. Changes became noticeable only at larger angles.

Figure~\ref{fig:DEcm} summarizes the error budget. Accordingly, the uncertainty of our collision energy scale amounts to $24 < \Delta E_\mathrm{cm} < 31$~meV for the  present range of electron-ion collision energies. In principle, uncertainties of the particle masses could also contribute to the error budget. However, these are below the ppm level and can thus safely be neglected. The partial derivatives required for error propagation have been straight-forwardly computed from Eq.~\ref{eq:Ecm}  and the individual contributions to $\Delta E_\mathrm{cm}$ by $\Delta E_d$, $\Delta E_\mathrm{cool}$, 
and $\Delta\theta$ have been added in quadrature. The largest contribution comes from the uncertainty in $E_d$, which is related to the long-term stability of the fast HV divider. In the future, we will use a newly-built more stable device, which has the potential to reduce the uncertainty of the voltage measurement by up to an order of magnitude. 

\begin{figure}[t]
	\centering{\includegraphics[width=\columnwidth]{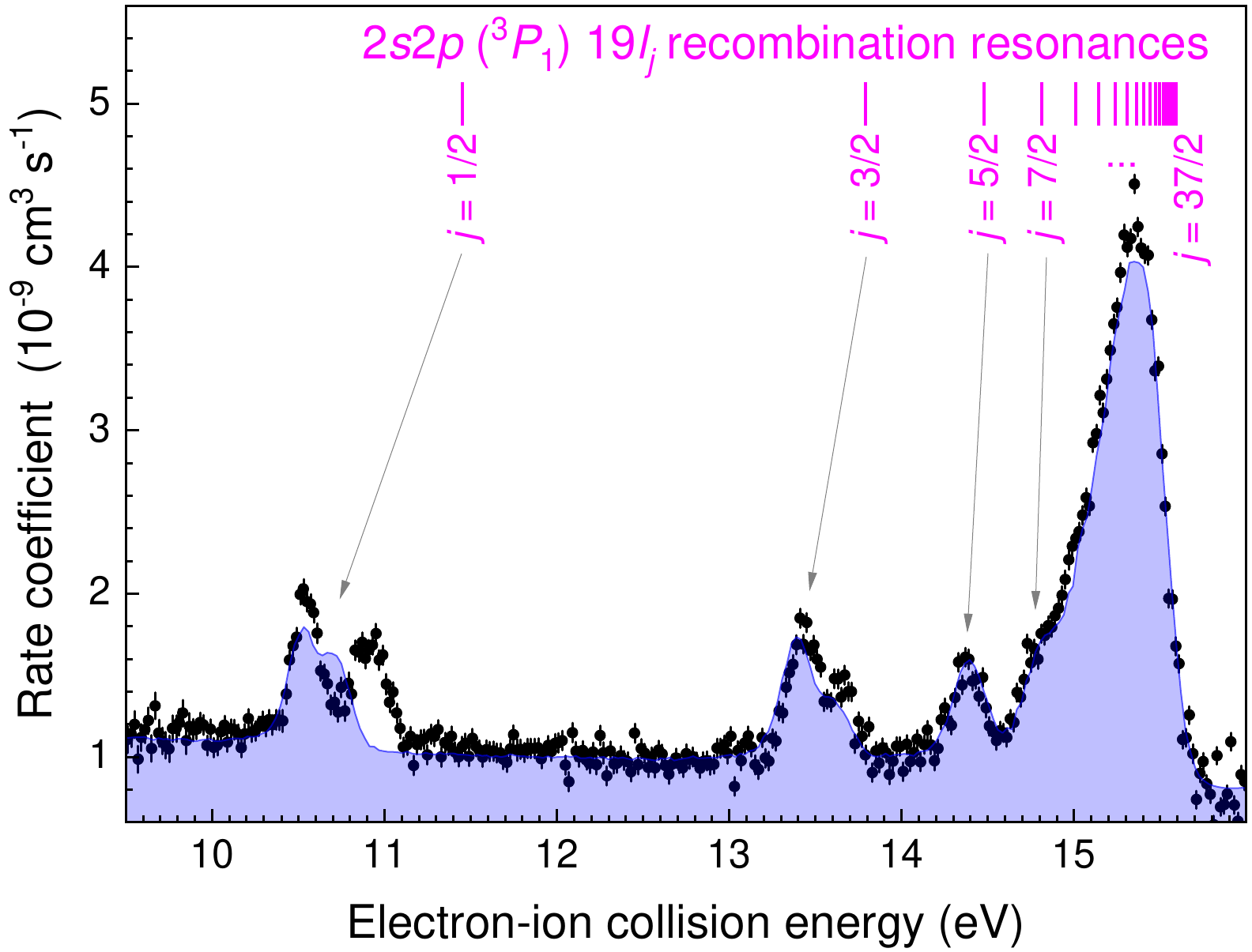}}
	\caption{\label{fig:ramp05}Comparison of the present experimental merged-beams rate coefficient for electron-ion recombination of Pb$^{78+}$ (symbols with statistical error bars) with the results of our theoretical calculations with the  \textsc{autostructure} code  \cite{Badnell2011b} (shaded full curve), where the $2s^2\;^1S_0 - 2s\,2p\;^3P_1$ excitation energy was adjusted to $E_\mathrm{exci}=244.937$~eV (see text). The vertical bars denote the $2s\,2p\,(^3P_1)\,19l_j$ DR resonance positions (Eq.~\eqref{eq:Eres}) using again $E_\mathrm{exci}=244.937$~eV and  for $E_b(nl_j)$  the Dirac formula for hydrogenlike ions with an effective nuclear charge of 78 for all 19 possible $j$ values ranging from $j=1/2$ to $j=19-1/2=37/2$ (see Ref.~\cite{Bernhardt2015a} for details).}
\end{figure}

Figure~\ref{fig:ramp05} shows our measured merged-beams rate coefficient for the recombination of Be-like Pb$^{78+}$ ions in the experimental  energy range 9.3--16.5~eV. In addition to the data from Fig.~\ref{fig:slowfast}, which were obtained with electrons both slower and faster than the ions, further experimental data sets contribute to Fig.~\ref{fig:ramp05}, which were measured with the electrons always being slower than the ions. The resonance features in the measured rate coefficient are associated with the $1s^2\,2s\,2p\,(^3P_{1})\,19l_j$ DR resonance levels, i.e., all belong to the $n=19$ manifold of Rydberg levels. Resonances associated with higher $n$ values would occur at higher energies beyond the current experimental energy range. In the experimental spectrum, the $j=1/2$ and $j=3/2$ resonances exhibit splittings into at least two components, which are due to the interaction  between the excited Be-like core and the Rydberg electron. The spacings between adjacent $j$ values become smaller for increasing $j$. At the given experimental resolving power, the resonance features associated with $7/2\leq j \leq 37/2$ could not be individually resolved.

An already quite reasonable estimate for DR resonance energies (Eq.~\ref{eq:Eres}, vertical bars in Fig.~\ref{fig:ramp05}) can be obtained by applying the Dirac formula for the $j$-dependent binding energies $E_b(19l_j)$ of the $n=19$ Ryd\-berg electron in a hydrogenlike ion with an effective charge of 78 and using  $E_\mathrm{exci}=244.937$~eV for the $1s^2\,2s^2\;^1S_0 - 1s^2\,2s\,2p\;^3P_1$ transition. The latter value has been derived from the present experiment (see below). The approach with the Dirac formula becomes increasingly better as $l$ (and, hence, $j$) increases. The reason for this is that the $l_j$ orbital becomes increasingly more hydrogenlike. The hydrogenic approximation also allows one to estimate the shifts of the Rydberg resonance energies in a plasma \cite{Iafrate1969}. At the experimental electron density of $10^7$~cm$^{-3}$ this \lq\lq{}continuum lowering\rq\rq\ is practically negligible (see appendix).

In order to obtain theoretical results also for the resonance strengths, which together with the resonance positions determine the shapes of the observed resonance features, we have performed DR calculations using the \textsc{autostructure} \cite{Badnell2011b} atomic code. Also in these calculations, we have adjusted the  $2s^2\;^1S_0 - 2s\,2p\;^3P_1$ excitation energy to 244.937~eV, which ensures that the theoretical result for the most prominent peak lines up with its experimental counterpart. 

In Fig.~\ref{fig:ramp05}, the theoretical results are compared with the experimental merged-beams rate coefficients. The theoretical calculations describe the experimental DR spectrum very well except for mismatches of the $j=1/2, 3/2$, and $5/2$ resonance groups due to deficiencies in handling correlation by \textsc{autostructure}. In particular, the strongest resonance feature, which is a blend of all $(2s\,2p\;^3P_1)\,19l_j$ resonances with $j\geq7/2$, agrees with the experimental peak position and height as was already previously achieved for DR of  Li-like Pb$^{79+}$ \cite{Badnell2004a}. The \textsc{autostructure} binding energies $E_b(19l_j)$ for $j\geq 7/2$ agree within 2~meV with the corresponding hydrogenic Dirac values. From this finding and from the fact that QED does not affect high-$j$ Rydberg resonance energies \cite{Shabaeva2003} we conclude that the uncertainty of $E_b(nl_j)$ in Eq.~\eqref{eq:Eres} is practically negligible.

Another potential contribution to the uncertainty of the peak positions arises from the comparison with the theoretical DR results. For the comparison the theoretical cross sections were convolved with the experimental collision-energy distribution. The latter was obtained from a Monte-Carlo approach similar to the one used previously for the ESR electron cooler \cite{Wang2024}. The Monte-Carlo convolution accounts for the longitudinal and transverse electron temperatures $kT_\|$ and $kT_\perp$, for the almost negligible ion temperature, and for the effects caused by the toroidal merging and demerging sections of the electron cooler \cite{Brandau2025}. In order to determine uncertainties resulting from the choices for the various parameters we performed systematic variations of $kT_\|$, $kT_\perp$, and $E_\mathrm{exci}$. From these we find that $kT_\|=0.23$~meV, $kT_\perp=3.3$~meV, and $E_\mathrm{exci}=244.937$~eV lead to the best match of theory and experiment in the energy range 14--16~eV of the $j\geq 5/2$ resonances (Fig.~\ref{fig:ramp05}).

The quoted temperature values correspond roughly to what is expected from the operation of the electron cooler with a beam expansion factor $\zeta=33$ \cite{Danared1994}. In particular, the transverse temperature is significantly lower than the one of the electron beam of the ESR electron cooler, where typically $kT_\perp=120$~meV \cite{Brandau2002a,Bernhardt2015a,Wang2024}, leading to a much reduced related uncertainty as compared to the experimental results from the ESR. From the above mentioned systematic variations we obtained a \lq{}fit\rq\ uncertainty of $\pm 3$~meV for $E_\mathrm{exci}$, which is negligible when added in quadrature to the $\pm 30$ systematic energy uncertainty. Thus, our experimental value for the Pb$^{78+}$($1s^2\,2s^2\;^1S_0 - 1s^2\,2s\,2p\;^3P_1$) excitation energy is $E_\mathrm{exci}^\mathrm{(exp)} = 244.937(30)$~eV.

Our experimental value agrees excellently with the theoretical value $E_\mathrm{exci}^\mathrm{(theo)} =244.942(52)$~eV of Malyshev et al.~\cite{Malyshev2024}. However, it significantly deviates from the earlier value 244.734~eV (no uncertainty estimate provided) by Cheng et~al.~\cite{Cheng2008a}. The largest contribution to the 210-meV difference between both theoretical values comes from the additional inclusion of all second-order QED contributions to the atomic binding energies by Malyshev et~al. With our present experimental uncertainty we test  the second-order strong-field QED calculations, which due to the near degeneracy of the atomic levels are much more challenging for a highly-correlated beryllium-like ion as compared to atomic systems with fewer electrons, on the 15\% level.  

In conclusion, we have successfully coupled a low-energy heavy-ion storage ring with a high-energy accelerator and storage-ring complex that is capable of providing few-electron ions of the heaviest elements at relatively low energies for precision electron-ion collision spectroscopy of a highly-correlated atomic system with an ultra-cold electron beam. Already the current uncertainty of the experimental energy scale challenges state-of-the-art atomic theory, since  our $\pm$30-meV experimental uncertainty is lower than the $\pm52$-meV theoretical one.

The future use of a recently developed high-voltage probe promises a significant increase of the experimental accuracy by up to an order of magnitude \cite{Brandau2025}. Our experimental accuracy will then be comparable to the presently most accurate results from x-ray spectroscopy \cite{Beiersdorfer2005a}. The statistical quality of the present experimental data was limited by unfavorable storage-ring vacuum conditions that led to a high recombination background due to charge capture in collisions of the stored ions with residual gas particles. The resulting low signal-to-background ratio forced us to operate the electron cooler with a relatively high electron density, by which the transverse electron temperature was compromised. In the future, we expect the residual-gas pressure in CRYRING  to be lower by an order of magnitude and, thus, less background, enhanced statistical data quality and higher experimental resolving power (using the maximum electron-beam expansion factor $\zeta=100$) than in the present pilot study. Our error analysis (Fig.~\ref{fig:DEcm}) suggests that, in general, higher experimental precision can be achieved for resonances that occur at lower electron-ion collision energies. According to the above mentioned estimates of DR resonance positions  U$^{88+}$ is among the heavy Be-like ions that feature DR resonances at energies below 10~eV. In future electron-ion collision experiments, we will use this heaviest ion to provide even more stringent constraints for theoretical predictions that involve the treatment of few-particle correlations and higher-order strong-field QED effects.

\begin{acknowledgements}

The results presented here are based on the experiment E131, which was performed at the heavy-ion storage ring CRYRING@ESR at the GSI Helmholtzzentrum f\"ur Schwerionenforschung, Darmstadt (Germany) in the frame of FAIR Phase-0. Financial support from the German Federal Ministry for Research, Technology, and Space (BMFTR) via the Collaborative Research Center ErUM-FSP T05 --- \lq\lq Aufbau von APPA bei FAIR\rq\rq\ (Grant Nos.\ 05P19PMFA1, 05P19RGFA1, 05P21PMFA1, 05P21RGFA1, 05P24RG2, and 05P24SJA) and from the National Key Research and Development Program of China under Grant No.\ 2022YFA1602500, and from the National Natural Science Foundation of China under Grant No.\ 12393824 is gratefully acknowledged. C.~B.\ and S.-X.~W.\ acknowledge the support by the State of Hesse within the Research Cluster ELEMENTS (Project ID 500/10.006). 

\end{acknowledgements}


\begin{thebibliography}{44}%
\makeatletter
\providecommand \@ifxundefined [1]{%
 \@ifx{#1\undefined}
}%
\providecommand \@ifnum [1]{%
 \ifnum #1\expandafter \@firstoftwo
 \else \expandafter \@secondoftwo
 \fi
}%
\providecommand \@ifx [1]{%
 \ifx #1\expandafter \@firstoftwo
 \else \expandafter \@secondoftwo
 \fi
}%
\providecommand \natexlab [1]{#1}%
\providecommand \enquote  [1]{``#1''}%
\providecommand \bibnamefont  [1]{#1}%
\providecommand \bibfnamefont [1]{#1}%
\providecommand \citenamefont [1]{#1}%
\providecommand \href@noop [0]{\@secondoftwo}%
\providecommand \href [0]{\begingroup \@sanitize@url \@href}%
\providecommand \@href[1]{\@@startlink{#1}\@@href}%
\providecommand \@@href[1]{\endgroup#1\@@endlink}%
\providecommand \@sanitize@url [0]{\catcode `\\12\catcode `\$12\catcode
  `\&12\catcode `\#12\catcode `\^12\catcode `\_12\catcode `\%12\relax}%
\providecommand \@@startlink[1]{}%
\providecommand \@@endlink[0]{}%
\providecommand \url  [0]{\begingroup\@sanitize@url \@url }%
\providecommand \@url [1]{\endgroup\@href {#1}{\urlprefix }}%
\providecommand \urlprefix  [0]{URL }%
\providecommand \Eprint [0]{\href }%
\providecommand \doibase [0]{https://doi.org/}%
\providecommand \selectlanguage [0]{\@gobble}%
\providecommand \bibinfo  [0]{\@secondoftwo}%
\providecommand \bibfield  [0]{\@secondoftwo}%
\providecommand \translation [1]{[#1]}%
\providecommand \BibitemOpen [0]{}%
\providecommand \bibitemStop [0]{}%
\providecommand \bibitemNoStop [0]{.\EOS\space}%
\providecommand \EOS [0]{\spacefactor3000\relax}%
\providecommand \BibitemShut  [1]{\csname bibitem#1\endcsname}%
\let\auto@bib@innerbib\@empty
\bibitem [{\citenamefont {Lindroth}\ \emph {et~al.}(2001)\citenamefont
  {Lindroth}, \citenamefont {Danared}, \citenamefont {Glans}, \citenamefont
  {Pesic}, \citenamefont {Tokman}, \citenamefont {Vikor},\ and\ \citenamefont
  {Schuch}}]{Lindroth2001}%
  \BibitemOpen
  \bibfield  {author} {\bibinfo {author} {\bibfnamefont {E.}~\bibnamefont
  {Lindroth}}, \bibinfo {author} {\bibfnamefont {H.}~\bibnamefont {Danared}},
  \bibinfo {author} {\bibfnamefont {P.}~\bibnamefont {Glans}}, \bibinfo
  {author} {\bibfnamefont {Z.}~\bibnamefont {Pesic}}, \bibinfo {author}
  {\bibfnamefont {M.}~\bibnamefont {Tokman}}, \bibinfo {author} {\bibfnamefont
  {G.}~\bibnamefont {Vikor}},\ and\ \bibinfo {author} {\bibfnamefont
  {R.}~\bibnamefont {Schuch}},\ }\bibfield  {title} {\bibinfo {title} {{QED}
  effects in {C}u-like {P}b recombination resonances near threshold},\ }\href
  {https://doi.org/10.1103/PhysRevLett.86.5027} {\bibfield  {journal} {\bibinfo
   {journal} {Phys. Rev. Lett.}\ }\textbf {\bibinfo {volume} {86}},\ \bibinfo
  {pages} {5027} (\bibinfo {year} {2001})}\BibitemShut {NoStop}%
\bibitem [{\citenamefont {Brandau}\ \emph {et~al.}(2003)\citenamefont
  {Brandau}, \citenamefont {Kozhuharov}, \citenamefont {M\"{u}ller},
  \citenamefont {Shi}, \citenamefont {Schippers}, \citenamefont {Bartsch},
  \citenamefont {B\"{o}hm}, \citenamefont {B\"{o}hme}, \citenamefont
  {Hoffknecht}, \citenamefont {Knopp}, \citenamefont {Gr\"{u}n}, \citenamefont
  {Scheid}, \citenamefont {Steih}, \citenamefont {Bosch}, \citenamefont
  {Franzke}, \citenamefont {Mokler}, \citenamefont {Nolden}, \citenamefont
  {Steck}, \citenamefont {St\"{o}hlker},\ and\ \citenamefont
  {Stachura}}]{Brandau2003b}%
  \BibitemOpen
  \bibfield  {author} {\bibinfo {author} {\bibfnamefont {C.}~\bibnamefont
  {Brandau}}, \bibinfo {author} {\bibfnamefont {C.}~\bibnamefont {Kozhuharov}},
  \bibinfo {author} {\bibfnamefont {A.}~\bibnamefont {M\"{u}ller}}, \bibinfo
  {author} {\bibfnamefont {W.}~\bibnamefont {Shi}}, \bibinfo {author}
  {\bibfnamefont {S.}~\bibnamefont {Schippers}}, \bibinfo {author}
  {\bibfnamefont {T.}~\bibnamefont {Bartsch}}, \bibinfo {author} {\bibfnamefont
  {S.}~\bibnamefont {B\"{o}hm}}, \bibinfo {author} {\bibfnamefont
  {C.}~\bibnamefont {B\"{o}hme}}, \bibinfo {author} {\bibfnamefont
  {A.}~\bibnamefont {Hoffknecht}}, \bibinfo {author} {\bibfnamefont
  {H.}~\bibnamefont {Knopp}}, \bibinfo {author} {\bibfnamefont
  {N.}~\bibnamefont {Gr\"{u}n}}, \bibinfo {author} {\bibfnamefont
  {W.}~\bibnamefont {Scheid}}, \bibinfo {author} {\bibfnamefont
  {T.}~\bibnamefont {Steih}}, \bibinfo {author} {\bibfnamefont
  {F.}~\bibnamefont {Bosch}}, \bibinfo {author} {\bibfnamefont
  {B.}~\bibnamefont {Franzke}}, \bibinfo {author} {\bibfnamefont {P.~H.}\
  \bibnamefont {Mokler}}, \bibinfo {author} {\bibfnamefont {F.}~\bibnamefont
  {Nolden}}, \bibinfo {author} {\bibfnamefont {M.}~\bibnamefont {Steck}},
  \bibinfo {author} {\bibfnamefont {T.}~\bibnamefont {St\"{o}hlker}},\ and\
  \bibinfo {author} {\bibfnamefont {Z.}~\bibnamefont {Stachura}},\ }\bibfield
  {title} {\bibinfo {title} {Precise determination of the $2s_{1/2} - 2p_{1/2}$
  splitting in very heavy lithiumlike ions utilizing dielectronic
  recombination},\ }\href {https://doi.org/10.1103/PhysRevLett.91.073202}
  {\bibfield  {journal} {\bibinfo  {journal} {Phys. Rev. Lett.}\ }\textbf
  {\bibinfo {volume} {91}},\ \bibinfo {pages} {073202} (\bibinfo {year}
  {2003})}\BibitemShut {NoStop}%
\bibitem [{\citenamefont {Lestinsky}\ \emph {et~al.}(2008)\citenamefont
  {Lestinsky}, \citenamefont {Lindroth}, \citenamefont {Orlov}, \citenamefont
  {Schmidt}, \citenamefont {Schippers}, \citenamefont {B\"{o}hm}, \citenamefont
  {Brandau}, \citenamefont {Sprenger}, \citenamefont {Terekhov}, \citenamefont
  {M\"{u}ller},\ and\ \citenamefont {Wolf}}]{Lestinsky2008a}%
  \BibitemOpen
  \bibfield  {author} {\bibinfo {author} {\bibfnamefont {M.}~\bibnamefont
  {Lestinsky}}, \bibinfo {author} {\bibfnamefont {E.}~\bibnamefont {Lindroth}},
  \bibinfo {author} {\bibfnamefont {D.~A.}\ \bibnamefont {Orlov}}, \bibinfo
  {author} {\bibfnamefont {E.~W.}\ \bibnamefont {Schmidt}}, \bibinfo {author}
  {\bibfnamefont {S.}~\bibnamefont {Schippers}}, \bibinfo {author}
  {\bibfnamefont {S.}~\bibnamefont {B\"{o}hm}}, \bibinfo {author}
  {\bibfnamefont {C.}~\bibnamefont {Brandau}}, \bibinfo {author} {\bibfnamefont
  {F.}~\bibnamefont {Sprenger}}, \bibinfo {author} {\bibfnamefont {A.~S.}\
  \bibnamefont {Terekhov}}, \bibinfo {author} {\bibfnamefont {A.}~\bibnamefont
  {M\"{u}ller}},\ and\ \bibinfo {author} {\bibfnamefont {A.}~\bibnamefont
  {Wolf}},\ }\bibfield  {title} {\bibinfo {title} {Screened radiative
  corrections from hyperfine-split dielectronic resonances in lithiumlike
  scandium},\ }\href {https://doi.org/10.1103/PhysRevLett.100.033001}
  {\bibfield  {journal} {\bibinfo  {journal} {Phys. Rev. Lett.}\ }\textbf
  {\bibinfo {volume} {100}},\ \bibinfo {pages} {033001} (\bibinfo {year}
  {2008})}\BibitemShut {NoStop}%
\bibitem [{\citenamefont {Schuch}\ \emph {et~al.}(2005)\citenamefont {Schuch},
  \citenamefont {Lindroth}, \citenamefont {Madzunkov}, \citenamefont {Fogle},
  \citenamefont {Mohamed},\ and\ \citenamefont {Indelicato}}]{Schuch2005a}%
  \BibitemOpen
  \bibfield  {author} {\bibinfo {author} {\bibfnamefont {R.}~\bibnamefont
  {Schuch}}, \bibinfo {author} {\bibfnamefont {E.}~\bibnamefont {Lindroth}},
  \bibinfo {author} {\bibfnamefont {S.}~\bibnamefont {Madzunkov}}, \bibinfo
  {author} {\bibfnamefont {M.}~\bibnamefont {Fogle}}, \bibinfo {author}
  {\bibfnamefont {T.}~\bibnamefont {Mohamed}},\ and\ \bibinfo {author}
  {\bibfnamefont {P.}~\bibnamefont {Indelicato}},\ }\bibfield  {title}
  {\bibinfo {title} {Dielectronic resonance method for measuring isotope
  shifts},\ }\href {https://doi.org/10.1103/PhysRevLett.95.183003} {\bibfield
  {journal} {\bibinfo  {journal} {Phys. Rev. Lett.}\ }\textbf {\bibinfo
  {volume} {95}},\ \bibinfo {pages} {183003} (\bibinfo {year}
  {2005})}\BibitemShut {NoStop}%
\bibitem [{\citenamefont {Brandau}\ \emph {et~al.}(2008)\citenamefont
  {Brandau}, \citenamefont {Kozhuharov}, \citenamefont {Harman}, \citenamefont
  {M\"{u}ller}, \citenamefont {Schippers}, \citenamefont {Kozhedub},
  \citenamefont {Bernhardt}, \citenamefont {B\"{o}hm}, \citenamefont {Jacobi},
  \citenamefont {Schmidt}, \citenamefont {Mokler}, \citenamefont {Bosch},
  \citenamefont {Kluge}, \citenamefont {St\"{o}hlker}, \citenamefont {Beckert},
  \citenamefont {Beller}, \citenamefont {Nolden}, \citenamefont {Steck},
  \citenamefont {Gumberidze}, \citenamefont {Reuschl}, \citenamefont
  {Spillmann}, \citenamefont {Currell}, \citenamefont {Tupitsyn}, \citenamefont
  {Shabaev}, \citenamefont {Jentschura}, \citenamefont {Keitel}, \citenamefont
  {Wolf},\ and\ \citenamefont {Stachura}}]{Brandau2008a}%
  \BibitemOpen
  \bibfield  {author} {\bibinfo {author} {\bibfnamefont {C.}~\bibnamefont
  {Brandau}}, \bibinfo {author} {\bibfnamefont {C.}~\bibnamefont {Kozhuharov}},
  \bibinfo {author} {\bibfnamefont {Z.}~\bibnamefont {Harman}}, \bibinfo
  {author} {\bibfnamefont {A.}~\bibnamefont {M\"{u}ller}}, \bibinfo {author}
  {\bibfnamefont {S.}~\bibnamefont {Schippers}}, \bibinfo {author}
  {\bibfnamefont {Y.~S.}\ \bibnamefont {Kozhedub}}, \bibinfo {author}
  {\bibfnamefont {D.}~\bibnamefont {Bernhardt}}, \bibinfo {author}
  {\bibfnamefont {S.}~\bibnamefont {B\"{o}hm}}, \bibinfo {author}
  {\bibfnamefont {J.}~\bibnamefont {Jacobi}}, \bibinfo {author} {\bibfnamefont
  {E.~W.}\ \bibnamefont {Schmidt}}, \bibinfo {author} {\bibfnamefont {P.~H.}\
  \bibnamefont {Mokler}}, \bibinfo {author} {\bibfnamefont {F.}~\bibnamefont
  {Bosch}}, \bibinfo {author} {\bibfnamefont {H.-J.}\ \bibnamefont {Kluge}},
  \bibinfo {author} {\bibfnamefont {T.}~\bibnamefont {St\"{o}hlker}}, \bibinfo
  {author} {\bibfnamefont {K.}~\bibnamefont {Beckert}}, \bibinfo {author}
  {\bibfnamefont {P.}~\bibnamefont {Beller}}, \bibinfo {author} {\bibfnamefont
  {F.}~\bibnamefont {Nolden}}, \bibinfo {author} {\bibfnamefont
  {M.}~\bibnamefont {Steck}}, \bibinfo {author} {\bibfnamefont
  {A.}~\bibnamefont {Gumberidze}}, \bibinfo {author} {\bibfnamefont
  {R.}~\bibnamefont {Reuschl}}, \bibinfo {author} {\bibfnamefont
  {U.}~\bibnamefont {Spillmann}}, \bibinfo {author} {\bibfnamefont {F.~J.}\
  \bibnamefont {Currell}}, \bibinfo {author} {\bibfnamefont {I.~I.}\
  \bibnamefont {Tupitsyn}}, \bibinfo {author} {\bibfnamefont {V.~M.}\
  \bibnamefont {Shabaev}}, \bibinfo {author} {\bibfnamefont {U.~D.}\
  \bibnamefont {Jentschura}}, \bibinfo {author} {\bibfnamefont {C.~H.}\
  \bibnamefont {Keitel}}, \bibinfo {author} {\bibfnamefont {A.}~\bibnamefont
  {Wolf}},\ and\ \bibinfo {author} {\bibfnamefont {Z.}~\bibnamefont
  {Stachura}},\ }\bibfield  {title} {\bibinfo {title} {Isotope shift in the
  dielectronic recombination of three-electron $^{A}${N}d$^{57+}$},\ }\href
  {https://doi.org/10.1103/PhysRevLett.100.073201} {\bibfield  {journal}
  {\bibinfo  {journal} {Phys. Rev. Lett.}\ }\textbf {\bibinfo {volume} {100}},\
  \bibinfo {pages} {073201} (\bibinfo {year} {2008})}\BibitemShut {NoStop}%
\bibitem [{\citenamefont {Schmidt}\ \emph {et~al.}(1994)\citenamefont
  {Schmidt}, \citenamefont {Forck}, \citenamefont {Grieser}, \citenamefont
  {Habs}, \citenamefont {Kenntner}, \citenamefont {Miersch}, \citenamefont
  {Repnow}, \citenamefont {Schramm}, \citenamefont {Sch\"{u}ssler},
  \citenamefont {Schwalm},\ and\ \citenamefont {Wolf}}]{Schmidt1994}%
  \BibitemOpen
  \bibfield  {author} {\bibinfo {author} {\bibfnamefont {H.~T.}\ \bibnamefont
  {Schmidt}}, \bibinfo {author} {\bibfnamefont {P.}~\bibnamefont {Forck}},
  \bibinfo {author} {\bibfnamefont {M.}~\bibnamefont {Grieser}}, \bibinfo
  {author} {\bibfnamefont {D.}~\bibnamefont {Habs}}, \bibinfo {author}
  {\bibfnamefont {J.}~\bibnamefont {Kenntner}}, \bibinfo {author}
  {\bibfnamefont {G.}~\bibnamefont {Miersch}}, \bibinfo {author} {\bibfnamefont
  {R.}~\bibnamefont {Repnow}}, \bibinfo {author} {\bibfnamefont
  {U.}~\bibnamefont {Schramm}}, \bibinfo {author} {\bibfnamefont
  {T.}~\bibnamefont {Sch\"{u}ssler}}, \bibinfo {author} {\bibfnamefont
  {D.}~\bibnamefont {Schwalm}},\ and\ \bibinfo {author} {\bibfnamefont
  {A.}~\bibnamefont {Wolf}},\ }\bibfield  {title} {\bibinfo {title}
  {High-precision measurement of the magnetic-dipole decay rate of metastable
  heliumlike carbon ions in a storage ring},\ }\href
  {https://doi.org/10.1103/PhysRevLett.72.1616} {\bibfield  {journal} {\bibinfo
   {journal} {Phys. Rev. Lett.}\ }\textbf {\bibinfo {volume} {72}},\ \bibinfo
  {pages} {1616} (\bibinfo {year} {1994})}\BibitemShut {NoStop}%
\bibitem [{\citenamefont {Schippers}\ \emph {et~al.}(2007)\citenamefont
  {Schippers}, \citenamefont {Schmidt}, \citenamefont {Bernhardt},
  \citenamefont {Yu}, \citenamefont {M\"{u}ller}, \citenamefont {Lestinsky},
  \citenamefont {Orlov}, \citenamefont {Grieser}, \citenamefont {Repnow},\ and\
  \citenamefont {Wolf}}]{Schippers2007a}%
  \BibitemOpen
  \bibfield  {author} {\bibinfo {author} {\bibfnamefont {S.}~\bibnamefont
  {Schippers}}, \bibinfo {author} {\bibfnamefont {E.~W.}\ \bibnamefont
  {Schmidt}}, \bibinfo {author} {\bibfnamefont {D.}~\bibnamefont {Bernhardt}},
  \bibinfo {author} {\bibfnamefont {D.}~\bibnamefont {Yu}}, \bibinfo {author}
  {\bibfnamefont {A.}~\bibnamefont {M\"{u}ller}}, \bibinfo {author}
  {\bibfnamefont {M.}~\bibnamefont {Lestinsky}}, \bibinfo {author}
  {\bibfnamefont {D.~A.}\ \bibnamefont {Orlov}}, \bibinfo {author}
  {\bibfnamefont {M.}~\bibnamefont {Grieser}}, \bibinfo {author} {\bibfnamefont
  {R.}~\bibnamefont {Repnow}},\ and\ \bibinfo {author} {\bibfnamefont
  {A.}~\bibnamefont {Wolf}},\ }\bibfield  {title} {\bibinfo {title}
  {Storage-ring measurement of the hyperfine induced \linebreak
  {$^{47}$Ti$^{18+}(2s\,2p\;^3P_0$} {$\to$} {$2s^2\;^1S_0)$} transition rate},\
  }\href {https://doi.org/10.1103/PhysRevLett.98.033001} {\bibfield  {journal}
  {\bibinfo  {journal} {Phys. Rev. Lett.}\ }\textbf {\bibinfo {volume} {98}},\
  \bibinfo {pages} {033001} (\bibinfo {year} {2007})}\BibitemShut {NoStop}%
\bibitem [{\citenamefont {Abrahamsson}\ \emph {et~al.}(1993)\citenamefont
  {Abrahamsson}, \citenamefont {Andler}, \citenamefont {Bagge}, \citenamefont
  {Beebe}, \citenamefont {Carl\'{e}}, \citenamefont {Danared}, \citenamefont
  {Egnell}, \citenamefont {Ehrnst\'{e}n}, \citenamefont {Engstr\"{o}m},
  \citenamefont {Herrlander}, \citenamefont {Hilke}, \citenamefont {Jeansson},
  \citenamefont {K\"{a}llberg}, \citenamefont {Leontein}, \citenamefont
  {Liljeby}, \citenamefont {Nilsson}, \citenamefont {Paal}, \citenamefont
  {Rensfelt}, \citenamefont {Roseng{\aa}rd}, \citenamefont {Simonsson},
  \citenamefont {Soltan}, \citenamefont {Starker},\ and\ \citenamefont
  {af~Ugglas}}]{Abrahamsson1993a}%
  \BibitemOpen
  \bibfield  {author} {\bibinfo {author} {\bibfnamefont {K.}~\bibnamefont
  {Abrahamsson}}, \bibinfo {author} {\bibfnamefont {G.}~\bibnamefont {Andler}},
  \bibinfo {author} {\bibfnamefont {L.}~\bibnamefont {Bagge}}, \bibinfo
  {author} {\bibfnamefont {E.}~\bibnamefont {Beebe}}, \bibinfo {author}
  {\bibfnamefont {P.}~\bibnamefont {Carl\'{e}}}, \bibinfo {author}
  {\bibfnamefont {H.}~\bibnamefont {Danared}}, \bibinfo {author} {\bibfnamefont
  {S.}~\bibnamefont {Egnell}}, \bibinfo {author} {\bibfnamefont
  {K.}~\bibnamefont {Ehrnst\'{e}n}}, \bibinfo {author} {\bibfnamefont
  {M.}~\bibnamefont {Engstr\"{o}m}}, \bibinfo {author} {\bibfnamefont {C.~J.}\
  \bibnamefont {Herrlander}}, \bibinfo {author} {\bibfnamefont
  {J.}~\bibnamefont {Hilke}}, \bibinfo {author} {\bibfnamefont
  {J.}~\bibnamefont {Jeansson}}, \bibinfo {author} {\bibfnamefont
  {A.}~\bibnamefont {K\"{a}llberg}}, \bibinfo {author} {\bibfnamefont
  {S.}~\bibnamefont {Leontein}}, \bibinfo {author} {\bibfnamefont
  {L.}~\bibnamefont {Liljeby}}, \bibinfo {author} {\bibfnamefont
  {A.}~\bibnamefont {Nilsson}}, \bibinfo {author} {\bibfnamefont
  {A.}~\bibnamefont {Paal}}, \bibinfo {author} {\bibfnamefont {K.-G.}\
  \bibnamefont {Rensfelt}}, \bibinfo {author} {\bibfnamefont {U.}~\bibnamefont
  {Roseng{\aa}rd}}, \bibinfo {author} {\bibfnamefont {A.}~\bibnamefont
  {Simonsson}}, \bibinfo {author} {\bibfnamefont {A.}~\bibnamefont {Soltan}},
  \bibinfo {author} {\bibfnamefont {J.}~\bibnamefont {Starker}},\ and\ \bibinfo
  {author} {\bibfnamefont {M.}~\bibnamefont {af~Ugglas}},\ }\bibfield  {title}
  {\bibinfo {title} {{CRYRING} --- a synchrotron, cooler and storage ring},\
  }\href {https://doi.org/10.1016/0168-583X(93)95341-2} {\bibfield  {journal}
  {\bibinfo  {journal} {Nucl. Instrum. Methods Phys. Res. B}\ }\textbf
  {\bibinfo {volume} {79}},\ \bibinfo {pages} {269} (\bibinfo {year}
  {1993})}\BibitemShut {NoStop}%
\bibitem [{\citenamefont {Danared}\ \emph {et~al.}(1994)\citenamefont
  {Danared}, \citenamefont {Andler}, \citenamefont {Bagge}, \citenamefont
  {Herrlander}, \citenamefont {Hilke}, \citenamefont {Jeansson}, \citenamefont
  {K\"{a}llberg}, \citenamefont {Nilsson}, \citenamefont {Pa\'{a}l},
  \citenamefont {Rensfelt}, \citenamefont {Roseng{\aa}rd}, \citenamefont
  {Starker},\ and\ \citenamefont {{af Ugglas}}}]{Danared1994}%
  \BibitemOpen
  \bibfield  {author} {\bibinfo {author} {\bibfnamefont {H.}~\bibnamefont
  {Danared}}, \bibinfo {author} {\bibfnamefont {G.}~\bibnamefont {Andler}},
  \bibinfo {author} {\bibfnamefont {L.}~\bibnamefont {Bagge}}, \bibinfo
  {author} {\bibfnamefont {C.~J.}\ \bibnamefont {Herrlander}}, \bibinfo
  {author} {\bibfnamefont {J.}~\bibnamefont {Hilke}}, \bibinfo {author}
  {\bibfnamefont {J.}~\bibnamefont {Jeansson}}, \bibinfo {author}
  {\bibfnamefont {A.}~\bibnamefont {K\"{a}llberg}}, \bibinfo {author}
  {\bibfnamefont {A.}~\bibnamefont {Nilsson}}, \bibinfo {author} {\bibfnamefont
  {A.}~\bibnamefont {Pa\'{a}l}}, \bibinfo {author} {\bibfnamefont {K.-G.}\
  \bibnamefont {Rensfelt}}, \bibinfo {author} {\bibfnamefont {U.}~\bibnamefont
  {Roseng{\aa}rd}}, \bibinfo {author} {\bibfnamefont {J.}~\bibnamefont
  {Starker}},\ and\ \bibinfo {author} {\bibfnamefont {M.}~\bibnamefont {{af
  Ugglas}}},\ }\bibfield  {title} {\bibinfo {title} {Electron cooling with an
  ultracold electron beam},\ }\href
  {https://doi.org/10.1103/PhysRevLett.72.3775} {\bibfield  {journal} {\bibinfo
   {journal} {Phys. Rev. Lett.}\ }\textbf {\bibinfo {volume} {72}},\ \bibinfo
  {pages} {3775} (\bibinfo {year} {1994})}\BibitemShut {NoStop}%
\bibitem [{\citenamefont {Madzunkov}\ \emph {et~al.}(2002)\citenamefont
  {Madzunkov}, \citenamefont {Lindroth}, \citenamefont {Ekl\"{o}w},
  \citenamefont {Tokman}, \citenamefont {Pa\'{a}l},\ and\ \citenamefont
  {Schuch}}]{Madzunkov2002}%
  \BibitemOpen
  \bibfield  {author} {\bibinfo {author} {\bibfnamefont {S.}~\bibnamefont
  {Madzunkov}}, \bibinfo {author} {\bibfnamefont {E.}~\bibnamefont {Lindroth}},
  \bibinfo {author} {\bibfnamefont {N.}~\bibnamefont {Ekl\"{o}w}}, \bibinfo
  {author} {\bibfnamefont {M.}~\bibnamefont {Tokman}}, \bibinfo {author}
  {\bibfnamefont {A.}~\bibnamefont {Pa\'{a}l}},\ and\ \bibinfo {author}
  {\bibfnamefont {R.}~\bibnamefont {Schuch}},\ }\bibfield  {title} {\bibinfo
  {title} {{QED} efffects in lithiumlike krypton},\ }\href
  {https://doi.org/10.1103/PhysRevA.65.032505} {\bibfield  {journal} {\bibinfo
  {journal} {Phys. Rev. A}\ }\textbf {\bibinfo {volume} {65}},\ \bibinfo
  {pages} {032505} (\bibinfo {year} {2002})}\BibitemShut {NoStop}%
\bibitem [{\citenamefont {Fogle}\ \emph {et~al.}(2003)\citenamefont {Fogle},
  \citenamefont {Ekl\"{o}w}, \citenamefont {Lindroth}, \citenamefont {Mohamed},
  \citenamefont {Schuch},\ and\ \citenamefont {Tokman}}]{Fogle2003a}%
  \BibitemOpen
  \bibfield  {author} {\bibinfo {author} {\bibfnamefont {M.}~\bibnamefont
  {Fogle}}, \bibinfo {author} {\bibfnamefont {N.}~\bibnamefont {Ekl\"{o}w}},
  \bibinfo {author} {\bibfnamefont {E.}~\bibnamefont {Lindroth}}, \bibinfo
  {author} {\bibfnamefont {T.}~\bibnamefont {Mohamed}}, \bibinfo {author}
  {\bibfnamefont {R.}~\bibnamefont {Schuch}},\ and\ \bibinfo {author}
  {\bibfnamefont {M.}~\bibnamefont {Tokman}},\ }\bibfield  {title} {\bibinfo
  {title} {Spectroscopic study of {M}g-like {N}i by means of dielectronic
  recombination of stored ions},\ }\href
  {https://doi.org/10.1088/0953-4075/36/12/314} {\bibfield  {journal} {\bibinfo
   {journal} {J. Phys. B}\ }\textbf {\bibinfo {volume} {36}},\ \bibinfo {pages}
  {2563} (\bibinfo {year} {2003})}\BibitemShut {NoStop}%
\bibitem [{\citenamefont {Lestinsky}\ \emph {et~al.}(2016)\citenamefont
  {Lestinsky}, \citenamefont {Andrianov}, \citenamefont {Aurand}, \citenamefont
  {Bagnoud}, \citenamefont {Bernhardt}, \citenamefont {Beyer}, \citenamefont
  {Bishop}, \citenamefont {Blaum}, \citenamefont {Bleile}, \citenamefont
  {Borovik}, \citenamefont {Bosch}, \citenamefont {Bostock}, \citenamefont
  {Brandau}, \citenamefont {Br\"{a}uning-Demian}, \citenamefont {Bray},
  \citenamefont {Davinson}, \citenamefont {Ebinger}, \citenamefont {Echler},
  \citenamefont {Egelhof}, \citenamefont {Ehresmann}, \citenamefont
  {Engstr\"{o}m}, \citenamefont {Enss}, \citenamefont {Ferreira}, \citenamefont
  {Fischer}, \citenamefont {Fleischmann}, \citenamefont {F\"{o}rster},
  \citenamefont {Fritzsche}, \citenamefont {Geithner}, \citenamefont {Geyer},
  \citenamefont {Glorius}, \citenamefont {G\"{o}bel}, \citenamefont {Gorda},
  \citenamefont {Goullon}, \citenamefont {Grabitz}, \citenamefont {Grisenti},
  \citenamefont {Gumberidze}, \citenamefont {Hagmann}, \citenamefont {Heil},
  \citenamefont {Heinz}, \citenamefont {Herfurth}, \citenamefont {He{\ss}},
  \citenamefont {Hillenbrand}, \citenamefont {Hubele}, \citenamefont
  {Indelicato}, \citenamefont {K\"{a}llberg}, \citenamefont {Kester},
  \citenamefont {Kiselev}, \citenamefont {Knie}, \citenamefont {Kozhuharov},
  \citenamefont {Kraft-Bermuth}, \citenamefont {K\"{u}hl}, \citenamefont
  {Lane}, \citenamefont {Litvinov}, \citenamefont {Liesen}, \citenamefont {Ma},
  \citenamefont {M\"{a}rtin}, \citenamefont {Moshammer}, \citenamefont
  {M\"{u}ller}, \citenamefont {Namba}, \citenamefont {Neumeyer}, \citenamefont
  {Nilsson}, \citenamefont {N\"{o}rtersh\"{a}user}, \citenamefont {Paulus},
  \citenamefont {Petridis}, \citenamefont {Reed}, \citenamefont {Reifarth},
  \citenamefont {Rei{\ss}}, \citenamefont {Rothhardt}, \citenamefont {Sanchez},
  \citenamefont {Sanjari}, \citenamefont {Schippers}, \citenamefont {Schmidt},
  \citenamefont {Schneider}, \citenamefont {Scholz}, \citenamefont {Schuch},
  \citenamefont {Schulz}, \citenamefont {Shabaev}, \citenamefont {Simonsson},
  \citenamefont {Sj\"{o}holm}, \citenamefont {Skeppstedt}, \citenamefont
  {Sonnabend}, \citenamefont {Spillmann}, \citenamefont {Stiebing},
  \citenamefont {Steck}, \citenamefont {St\"{o}hlker}, \citenamefont
  {Surzhykov}, \citenamefont {Torilov}, \citenamefont {Tr\"{a}bert},
  \citenamefont {Trassinelli}, \citenamefont {Trotsenko}, \citenamefont {Tu},
  \citenamefont {Uschmann}, \citenamefont {Walker}, \citenamefont {Weber},
  \citenamefont {Winters}, \citenamefont {Woods}, \citenamefont {Zhao},\ and\
  \citenamefont {Zhang}}]{Lestinsky2016}%
  \BibitemOpen
  \bibfield  {author} {\bibinfo {author} {\bibfnamefont {M.}~\bibnamefont
  {Lestinsky}}, \bibinfo {author} {\bibfnamefont {V.}~\bibnamefont
  {Andrianov}}, \bibinfo {author} {\bibfnamefont {B.}~\bibnamefont {Aurand}},
  \bibinfo {author} {\bibfnamefont {V.}~\bibnamefont {Bagnoud}}, \bibinfo
  {author} {\bibfnamefont {D.}~\bibnamefont {Bernhardt}}, \bibinfo {author}
  {\bibfnamefont {H.}~\bibnamefont {Beyer}}, \bibinfo {author} {\bibfnamefont
  {S.}~\bibnamefont {Bishop}}, \bibinfo {author} {\bibfnamefont
  {K.}~\bibnamefont {Blaum}}, \bibinfo {author} {\bibfnamefont
  {A.}~\bibnamefont {Bleile}}, \bibinfo {author} {\bibfnamefont
  {A.}~\bibnamefont {Borovik}}, \bibinfo {author} {\bibfnamefont
  {F.}~\bibnamefont {Bosch}}, \bibinfo {author} {\bibfnamefont {C.~J.}\
  \bibnamefont {Bostock}}, \bibinfo {author} {\bibfnamefont {C.}~\bibnamefont
  {Brandau}}, \bibinfo {author} {\bibfnamefont {A.}~\bibnamefont
  {Br\"{a}uning-Demian}}, \bibinfo {author} {\bibfnamefont {I.}~\bibnamefont
  {Bray}}, \bibinfo {author} {\bibfnamefont {T.}~\bibnamefont {Davinson}},
  \bibinfo {author} {\bibfnamefont {B.}~\bibnamefont {Ebinger}}, \bibinfo
  {author} {\bibfnamefont {A.}~\bibnamefont {Echler}}, \bibinfo {author}
  {\bibfnamefont {P.}~\bibnamefont {Egelhof}}, \bibinfo {author} {\bibfnamefont
  {A.}~\bibnamefont {Ehresmann}}, \bibinfo {author} {\bibfnamefont
  {M.}~\bibnamefont {Engstr\"{o}m}}, \bibinfo {author} {\bibfnamefont
  {C.}~\bibnamefont {Enss}}, \bibinfo {author} {\bibfnamefont {N.}~\bibnamefont
  {Ferreira}}, \bibinfo {author} {\bibfnamefont {D.}~\bibnamefont {Fischer}},
  \bibinfo {author} {\bibfnamefont {A.}~\bibnamefont {Fleischmann}}, \bibinfo
  {author} {\bibfnamefont {E.}~\bibnamefont {F\"{o}rster}}, \bibinfo {author}
  {\bibfnamefont {S.}~\bibnamefont {Fritzsche}}, \bibinfo {author}
  {\bibfnamefont {R.}~\bibnamefont {Geithner}}, \bibinfo {author}
  {\bibfnamefont {S.}~\bibnamefont {Geyer}}, \bibinfo {author} {\bibfnamefont
  {J.}~\bibnamefont {Glorius}}, \bibinfo {author} {\bibfnamefont
  {K.}~\bibnamefont {G\"{o}bel}}, \bibinfo {author} {\bibfnamefont
  {O.}~\bibnamefont {Gorda}}, \bibinfo {author} {\bibfnamefont
  {J.}~\bibnamefont {Goullon}}, \bibinfo {author} {\bibfnamefont
  {P.}~\bibnamefont {Grabitz}}, \bibinfo {author} {\bibfnamefont
  {R.}~\bibnamefont {Grisenti}}, \bibinfo {author} {\bibfnamefont
  {A.}~\bibnamefont {Gumberidze}}, \bibinfo {author} {\bibfnamefont
  {S.}~\bibnamefont {Hagmann}}, \bibinfo {author} {\bibfnamefont
  {M.}~\bibnamefont {Heil}}, \bibinfo {author} {\bibfnamefont {A.}~\bibnamefont
  {Heinz}}, \bibinfo {author} {\bibfnamefont {F.}~\bibnamefont {Herfurth}},
  \bibinfo {author} {\bibfnamefont {R.}~\bibnamefont {He{\ss}}}, \bibinfo
  {author} {\bibfnamefont {P.-M.}\ \bibnamefont {Hillenbrand}}, \bibinfo
  {author} {\bibfnamefont {R.}~\bibnamefont {Hubele}}, \bibinfo {author}
  {\bibfnamefont {P.}~\bibnamefont {Indelicato}}, \bibinfo {author}
  {\bibfnamefont {A.}~\bibnamefont {K\"{a}llberg}}, \bibinfo {author}
  {\bibfnamefont {O.}~\bibnamefont {Kester}}, \bibinfo {author} {\bibfnamefont
  {O.}~\bibnamefont {Kiselev}}, \bibinfo {author} {\bibfnamefont
  {A.}~\bibnamefont {Knie}}, \bibinfo {author} {\bibfnamefont {C.}~\bibnamefont
  {Kozhuharov}}, \bibinfo {author} {\bibfnamefont {S.}~\bibnamefont
  {Kraft-Bermuth}}, \bibinfo {author} {\bibfnamefont {T.}~\bibnamefont
  {K\"{u}hl}}, \bibinfo {author} {\bibfnamefont {G.}~\bibnamefont {Lane}},
  \bibinfo {author} {\bibfnamefont {Y.~A.}\ \bibnamefont {Litvinov}}, \bibinfo
  {author} {\bibfnamefont {D.}~\bibnamefont {Liesen}}, \bibinfo {author}
  {\bibfnamefont {X.~W.}\ \bibnamefont {Ma}}, \bibinfo {author} {\bibfnamefont
  {R.}~\bibnamefont {M\"{a}rtin}}, \bibinfo {author} {\bibfnamefont
  {R.}~\bibnamefont {Moshammer}}, \bibinfo {author} {\bibfnamefont
  {A.}~\bibnamefont {M\"{u}ller}}, \bibinfo {author} {\bibfnamefont
  {S.}~\bibnamefont {Namba}}, \bibinfo {author} {\bibfnamefont
  {P.}~\bibnamefont {Neumeyer}}, \bibinfo {author} {\bibfnamefont
  {T.}~\bibnamefont {Nilsson}}, \bibinfo {author} {\bibfnamefont
  {W.}~\bibnamefont {N\"{o}rtersh\"{a}user}}, \bibinfo {author} {\bibfnamefont
  {G.}~\bibnamefont {Paulus}}, \bibinfo {author} {\bibfnamefont
  {N.}~\bibnamefont {Petridis}}, \bibinfo {author} {\bibfnamefont
  {M.}~\bibnamefont {Reed}}, \bibinfo {author} {\bibfnamefont {R.}~\bibnamefont
  {Reifarth}}, \bibinfo {author} {\bibfnamefont {P.}~\bibnamefont {Rei{\ss}}},
  \bibinfo {author} {\bibfnamefont {J.}~\bibnamefont {Rothhardt}}, \bibinfo
  {author} {\bibfnamefont {R.}~\bibnamefont {Sanchez}}, \bibinfo {author}
  {\bibfnamefont {M.~S.}\ \bibnamefont {Sanjari}}, \bibinfo {author}
  {\bibfnamefont {S.}~\bibnamefont {Schippers}}, \bibinfo {author}
  {\bibfnamefont {H.~T.}\ \bibnamefont {Schmidt}}, \bibinfo {author}
  {\bibfnamefont {D.}~\bibnamefont {Schneider}}, \bibinfo {author}
  {\bibfnamefont {P.}~\bibnamefont {Scholz}}, \bibinfo {author} {\bibfnamefont
  {R.}~\bibnamefont {Schuch}}, \bibinfo {author} {\bibfnamefont
  {M.}~\bibnamefont {Schulz}}, \bibinfo {author} {\bibfnamefont
  {V.}~\bibnamefont {Shabaev}}, \bibinfo {author} {\bibfnamefont
  {A.}~\bibnamefont {Simonsson}}, \bibinfo {author} {\bibfnamefont
  {J.}~\bibnamefont {Sj\"{o}holm}}, \bibinfo {author} {\bibfnamefont
  {O.}~\bibnamefont {Skeppstedt}}, \bibinfo {author} {\bibfnamefont
  {K.}~\bibnamefont {Sonnabend}}, \bibinfo {author} {\bibfnamefont
  {U.}~\bibnamefont {Spillmann}}, \bibinfo {author} {\bibfnamefont
  {K.}~\bibnamefont {Stiebing}}, \bibinfo {author} {\bibfnamefont
  {M.}~\bibnamefont {Steck}}, \bibinfo {author} {\bibfnamefont
  {T.}~\bibnamefont {St\"{o}hlker}}, \bibinfo {author} {\bibfnamefont
  {A.}~\bibnamefont {Surzhykov}}, \bibinfo {author} {\bibfnamefont
  {S.}~\bibnamefont {Torilov}}, \bibinfo {author} {\bibfnamefont
  {E.}~\bibnamefont {Tr\"{a}bert}}, \bibinfo {author} {\bibfnamefont
  {M.}~\bibnamefont {Trassinelli}}, \bibinfo {author} {\bibfnamefont
  {S.}~\bibnamefont {Trotsenko}}, \bibinfo {author} {\bibfnamefont {X.~L.}\
  \bibnamefont {Tu}}, \bibinfo {author} {\bibfnamefont {I.}~\bibnamefont
  {Uschmann}}, \bibinfo {author} {\bibfnamefont {P.~M.}\ \bibnamefont
  {Walker}}, \bibinfo {author} {\bibfnamefont {G.}~\bibnamefont {Weber}},
  \bibinfo {author} {\bibfnamefont {D.~F.~A.}\ \bibnamefont {Winters}},
  \bibinfo {author} {\bibfnamefont {P.~J.}\ \bibnamefont {Woods}}, \bibinfo
  {author} {\bibfnamefont {H.~Y.}\ \bibnamefont {Zhao}},\ and\ \bibinfo
  {author} {\bibfnamefont {Y.~H.}\ \bibnamefont {Zhang}},\ }\bibfield  {title}
  {\bibinfo {title} {Physics book: {CRYRING@ESR}},\ }\href
  {https://doi.org/10.1140/epjst/e2016-02643-6} {\bibfield  {journal} {\bibinfo
   {journal} {Eur. Phys. J. ST}\ }\textbf {\bibinfo {volume} {225}},\ \bibinfo
  {pages} {797} (\bibinfo {year} {2016})}\BibitemShut {NoStop}%
\bibitem [{\citenamefont {Zhu}\ \emph {et~al.}(2022)\citenamefont {Zhu},
  \citenamefont {Gumberidze}, \citenamefont {Over}, \citenamefont {Weber},
  \citenamefont {Andelkovic}, \citenamefont {Br\"auning-Demian}, \citenamefont
  {Chen}, \citenamefont {Dmytriiev}, \citenamefont {Forstner}, \citenamefont
  {Hahn}, \citenamefont {Herfurth}, \citenamefont {Herdrich}, \citenamefont
  {Hillenbrand}, \citenamefont {Kalinin}, \citenamefont {Kr\"oger},
  \citenamefont {Lestinsky}, \citenamefont {Litvinov}, \citenamefont {Menz},
  \citenamefont {Middents}, \citenamefont {Morgenroth}, \citenamefont
  {Petridis}, \citenamefont {Pf\"afflein}, \citenamefont {Sanjari},
  \citenamefont {Sidhu}, \citenamefont {Spillmann}, \citenamefont {Schuch},
  \citenamefont {Schippers}, \citenamefont {Trotsenko}, \citenamefont {Varga},
  \citenamefont {Vorobyev},\ and\ \citenamefont {St\"ohlker}}]{Zhu2022}%
  \BibitemOpen
  \bibfield  {author} {\bibinfo {author} {\bibfnamefont {B.}~\bibnamefont
  {Zhu}}, \bibinfo {author} {\bibfnamefont {A.}~\bibnamefont {Gumberidze}},
  \bibinfo {author} {\bibfnamefont {T.}~\bibnamefont {Over}}, \bibinfo {author}
  {\bibfnamefont {G.}~\bibnamefont {Weber}}, \bibinfo {author} {\bibfnamefont
  {Z.}~\bibnamefont {Andelkovic}}, \bibinfo {author} {\bibfnamefont
  {A.}~\bibnamefont {Br\"auning-Demian}}, \bibinfo {author} {\bibfnamefont
  {R.~J.}\ \bibnamefont {Chen}}, \bibinfo {author} {\bibfnamefont
  {D.}~\bibnamefont {Dmytriiev}}, \bibinfo {author} {\bibfnamefont
  {O.}~\bibnamefont {Forstner}}, \bibinfo {author} {\bibfnamefont
  {C.}~\bibnamefont {Hahn}}, \bibinfo {author} {\bibfnamefont {F.}~\bibnamefont
  {Herfurth}}, \bibinfo {author} {\bibfnamefont {M.~O.}\ \bibnamefont
  {Herdrich}}, \bibinfo {author} {\bibfnamefont {P.-M.}\ \bibnamefont
  {Hillenbrand}}, \bibinfo {author} {\bibfnamefont {A.}~\bibnamefont
  {Kalinin}}, \bibinfo {author} {\bibfnamefont {F.~M.}\ \bibnamefont
  {Kr\"oger}}, \bibinfo {author} {\bibfnamefont {M.}~\bibnamefont {Lestinsky}},
  \bibinfo {author} {\bibfnamefont {Y.~A.}\ \bibnamefont {Litvinov}}, \bibinfo
  {author} {\bibfnamefont {E.~B.}\ \bibnamefont {Menz}}, \bibinfo {author}
  {\bibfnamefont {W.}~\bibnamefont {Middents}}, \bibinfo {author}
  {\bibfnamefont {T.}~\bibnamefont {Morgenroth}}, \bibinfo {author}
  {\bibfnamefont {N.}~\bibnamefont {Petridis}}, \bibinfo {author}
  {\bibfnamefont {P.}~\bibnamefont {Pf\"afflein}}, \bibinfo {author}
  {\bibfnamefont {M.~S.}\ \bibnamefont {Sanjari}}, \bibinfo {author}
  {\bibfnamefont {R.~S.}\ \bibnamefont {Sidhu}}, \bibinfo {author}
  {\bibfnamefont {U.}~\bibnamefont {Spillmann}}, \bibinfo {author}
  {\bibfnamefont {R.}~\bibnamefont {Schuch}}, \bibinfo {author} {\bibfnamefont
  {S.}~\bibnamefont {Schippers}}, \bibinfo {author} {\bibfnamefont
  {S.}~\bibnamefont {Trotsenko}}, \bibinfo {author} {\bibfnamefont
  {L.}~\bibnamefont {Varga}}, \bibinfo {author} {\bibfnamefont
  {G.}~\bibnamefont {Vorobyev}},\ and\ \bibinfo {author} {\bibfnamefont
  {T.}~\bibnamefont {St\"ohlker}},\ }\bibfield  {title} {\bibinfo {title}
  {X-ray emission associated with radiative recombination for {Pb$^{82+}$} ions
  at threshold energies},\ }\href {https://doi.org/10.1103/PhysRevA.105.052804}
  {\bibfield  {journal} {\bibinfo  {journal} {Phys. Rev. A}\ }\textbf {\bibinfo
  {volume} {105}},\ \bibinfo {pages} {052804} (\bibinfo {year}
  {2022})}\BibitemShut {NoStop}%
\bibitem [{\citenamefont {Brandau}\ \emph {et~al.}(2025)\citenamefont
  {Brandau}, \citenamefont {Fuchs}, \citenamefont {Hannen}, \citenamefont
  {Hanu}, \citenamefont {Krantz}, \citenamefont {Lestinsky}, \citenamefont
  {Looshorn}, \citenamefont {Menz}, \citenamefont {Wang},\ and\ \citenamefont
  {Schippers}}]{Brandau2025}%
  \BibitemOpen
  \bibfield  {author} {\bibinfo {author} {\bibfnamefont {C.}~\bibnamefont
  {Brandau}}, \bibinfo {author} {\bibfnamefont {S.}~\bibnamefont {Fuchs}},
  \bibinfo {author} {\bibfnamefont {V.}~\bibnamefont {Hannen}}, \bibinfo
  {author} {\bibfnamefont {E.~O.}\ \bibnamefont {Hanu}}, \bibinfo {author}
  {\bibfnamefont {C.}~\bibnamefont {Krantz}}, \bibinfo {author} {\bibfnamefont
  {M.}~\bibnamefont {Lestinsky}}, \bibinfo {author} {\bibfnamefont
  {M.}~\bibnamefont {Looshorn}}, \bibinfo {author} {\bibfnamefont {E.~B.}\
  \bibnamefont {Menz}}, \bibinfo {author} {\bibfnamefont {S.~X.}\ \bibnamefont
  {Wang}},\ and\ \bibinfo {author} {\bibfnamefont {S.}~\bibnamefont
  {Schippers}},\ }\bibfield  {title} {\bibinfo {title} {Electron-ion collision
  spectroscopy at the {CRYRING@ESR} electron cooler},\ }\href
  {https://doi.org/10.1088/1674-1137/adbf81} {\bibfield  {journal} {\bibinfo
  {journal} {Chin. Phys. C}\ }\textbf {\bibinfo {volume} {49}},\ \bibinfo
  {pages} {064001} (\bibinfo {year} {2025})}\BibitemShut {NoStop}%
\bibitem [{\citenamefont {Malyshev}\ \emph {et~al.}(2021)\citenamefont
  {Malyshev}, \citenamefont {Glazov}, \citenamefont {Kozhedub}, \citenamefont
  {Anisimova}, \citenamefont {Kaygorodov}, \citenamefont {Shabaev},\ and\
  \citenamefont {Tupitsyn}}]{Malyshev2021}%
  \BibitemOpen
  \bibfield  {author} {\bibinfo {author} {\bibfnamefont {A.~V.}\ \bibnamefont
  {Malyshev}}, \bibinfo {author} {\bibfnamefont {D.~A.}\ \bibnamefont
  {Glazov}}, \bibinfo {author} {\bibfnamefont {Y.~S.}\ \bibnamefont
  {Kozhedub}}, \bibinfo {author} {\bibfnamefont {I.~S.}\ \bibnamefont
  {Anisimova}}, \bibinfo {author} {\bibfnamefont {M.~Y.}\ \bibnamefont
  {Kaygorodov}}, \bibinfo {author} {\bibfnamefont {V.~M.}\ \bibnamefont
  {Shabaev}},\ and\ \bibinfo {author} {\bibfnamefont {I.~I.}\ \bibnamefont
  {Tupitsyn}},\ }\bibfield  {title} {\bibinfo {title} {Ab initio calculations
  of energy levels in {B}e-like xenon: strong interference between
  electron-correlation and {QED} effects},\ }\href
  {https://doi.org/10.1103/PhysRevLett.126.183001} {\bibfield  {journal}
  {\bibinfo  {journal} {Phys. Rev. Lett.}\ }\textbf {\bibinfo {volume} {126}},\
  \bibinfo {pages} {183001} (\bibinfo {year} {2021})}\BibitemShut {NoStop}%
\bibitem [{\citenamefont {Beiersdorfer}\ \emph {et~al.}(1993)\citenamefont
  {Beiersdorfer}, \citenamefont {Knapp}, \citenamefont {Marrs}, \citenamefont
  {Elliott},\ and\ \citenamefont {Chen}}]{Beiersdorfer1993}%
  \BibitemOpen
  \bibfield  {author} {\bibinfo {author} {\bibfnamefont {P.}~\bibnamefont
  {Beiersdorfer}}, \bibinfo {author} {\bibfnamefont {D.}~\bibnamefont {Knapp}},
  \bibinfo {author} {\bibfnamefont {R.~E.}\ \bibnamefont {Marrs}}, \bibinfo
  {author} {\bibfnamefont {S.~R.}\ \bibnamefont {Elliott}},\ and\ \bibinfo
  {author} {\bibfnamefont {M.~H.}\ \bibnamefont {Chen}},\ }\bibfield  {title}
  {\bibinfo {title} {Structure and {L}amb shift of {$2s_{1/2} - 2p_{3/2}$}
  levels in lithiumlike {U$^{\,89+}$} through neonlike {U$^{\,82+}$}},\ }\href
  {https://doi.org/10.1103/PhysRevLett.71.3939} {\bibfield  {journal} {\bibinfo
   {journal} {Phys. Rev. Lett.}\ }\textbf {\bibinfo {volume} {71}},\ \bibinfo
  {pages} {3939} (\bibinfo {year} {1993})}\BibitemShut {NoStop}%
\bibitem [{\citenamefont {Beiersdorfer}\ \emph {et~al.}(2005)\citenamefont
  {Beiersdorfer}, \citenamefont {Chen}, \citenamefont {Thorn},\ and\
  \citenamefont {Tr\"{a}bert}}]{Beiersdorfer2005a}%
  \BibitemOpen
  \bibfield  {author} {\bibinfo {author} {\bibfnamefont {P.}~\bibnamefont
  {Beiersdorfer}}, \bibinfo {author} {\bibfnamefont {H.}~\bibnamefont {Chen}},
  \bibinfo {author} {\bibfnamefont {D.~B.}\ \bibnamefont {Thorn}},\ and\
  \bibinfo {author} {\bibfnamefont {E.}~\bibnamefont {Tr\"{a}bert}},\
  }\bibfield  {title} {\bibinfo {title} {Measurement of the two-loop {L}amb
  shift in lithiumlike {U}$^{89+}$},\ }\href
  {https://doi.org/10.1103/PhysRevLett.95.233003} {\bibfield  {journal}
  {\bibinfo  {journal} {Phys. Rev. Lett.}\ }\textbf {\bibinfo {volume} {95}},\
  \bibinfo {pages} {233003} (\bibinfo {year} {2005})}\BibitemShut {NoStop}%
\bibitem [{\citenamefont {Loetzsch}\ \emph {et~al.}(2024)\citenamefont
  {Loetzsch}, \citenamefont {Beyer}, \citenamefont {Duval}, \citenamefont
  {Spillmann}, \citenamefont {Bana{{\'{s}}}}, \citenamefont {Dergham},
  \citenamefont {Kr{\"{o}}ger}, \citenamefont {Glorius}, \citenamefont
  {Grisenti}, \citenamefont {Guerra}, \citenamefont {Gumberidze}, \citenamefont
  {He{\ss}}, \citenamefont {Hillenbrand}, \citenamefont {Indelicato},
  \citenamefont {Jagodzinski}, \citenamefont {Lamour}, \citenamefont {Lorentz},
  \citenamefont {Litvinov}, \citenamefont {Litvinov}, \citenamefont {Machado},
  \citenamefont {Paul}, \citenamefont {Paulus}, \citenamefont {Petridis},
  \citenamefont {Santos}, \citenamefont {Scheidel}, \citenamefont {Sidhu},
  \citenamefont {Steck}, \citenamefont {Steydli}, \citenamefont {Szary},
  \citenamefont {Trotsenko}, \citenamefont {Uschmann}, \citenamefont {Weber},
  \citenamefont {St{\"{o}}hlker},\ and\ \citenamefont
  {Trassinelli}}]{Loetzsch2024}%
  \BibitemOpen
  \bibfield  {author} {\bibinfo {author} {\bibfnamefont {R.}~\bibnamefont
  {Loetzsch}}, \bibinfo {author} {\bibfnamefont {H.~F.}\ \bibnamefont {Beyer}},
  \bibinfo {author} {\bibfnamefont {L.}~\bibnamefont {Duval}}, \bibinfo
  {author} {\bibfnamefont {U.}~\bibnamefont {Spillmann}}, \bibinfo {author}
  {\bibfnamefont {D.}~\bibnamefont {Bana{{\'{s}}}}}, \bibinfo {author}
  {\bibfnamefont {P.}~\bibnamefont {Dergham}}, \bibinfo {author} {\bibfnamefont
  {F.~M.}\ \bibnamefont {Kr{\"{o}}ger}}, \bibinfo {author} {\bibfnamefont
  {J.}~\bibnamefont {Glorius}}, \bibinfo {author} {\bibfnamefont {R.~E.}\
  \bibnamefont {Grisenti}}, \bibinfo {author} {\bibfnamefont {M.}~\bibnamefont
  {Guerra}}, \bibinfo {author} {\bibfnamefont {A.}~\bibnamefont {Gumberidze}},
  \bibinfo {author} {\bibfnamefont {R.}~\bibnamefont {He{\ss}}}, \bibinfo
  {author} {\bibfnamefont {P.-M.}\ \bibnamefont {Hillenbrand}}, \bibinfo
  {author} {\bibfnamefont {P.}~\bibnamefont {Indelicato}}, \bibinfo {author}
  {\bibfnamefont {P.}~\bibnamefont {Jagodzinski}}, \bibinfo {author}
  {\bibfnamefont {E.}~\bibnamefont {Lamour}}, \bibinfo {author} {\bibfnamefont
  {B.}~\bibnamefont {Lorentz}}, \bibinfo {author} {\bibfnamefont
  {S.}~\bibnamefont {Litvinov}}, \bibinfo {author} {\bibfnamefont {Y.~A.}\
  \bibnamefont {Litvinov}}, \bibinfo {author} {\bibfnamefont {J.}~\bibnamefont
  {Machado}}, \bibinfo {author} {\bibfnamefont {N.}~\bibnamefont {Paul}},
  \bibinfo {author} {\bibfnamefont {G.~G.}\ \bibnamefont {Paulus}}, \bibinfo
  {author} {\bibfnamefont {N.}~\bibnamefont {Petridis}}, \bibinfo {author}
  {\bibfnamefont {J.~P.}\ \bibnamefont {Santos}}, \bibinfo {author}
  {\bibfnamefont {M.}~\bibnamefont {Scheidel}}, \bibinfo {author}
  {\bibfnamefont {R.~S.}\ \bibnamefont {Sidhu}}, \bibinfo {author}
  {\bibfnamefont {M.}~\bibnamefont {Steck}}, \bibinfo {author} {\bibfnamefont
  {S.}~\bibnamefont {Steydli}}, \bibinfo {author} {\bibfnamefont
  {K.}~\bibnamefont {Szary}}, \bibinfo {author} {\bibfnamefont
  {S.}~\bibnamefont {Trotsenko}}, \bibinfo {author} {\bibfnamefont
  {I.}~\bibnamefont {Uschmann}}, \bibinfo {author} {\bibfnamefont
  {G.}~\bibnamefont {Weber}}, \bibinfo {author} {\bibfnamefont
  {T.}~\bibnamefont {St{\"{o}}hlker}},\ and\ \bibinfo {author} {\bibfnamefont
  {M.}~\bibnamefont {Trassinelli}},\ }\bibfield  {title} {\bibinfo {title}
  {Testing quantum electrodynamics in extreme fields using helium-like
  uranium},\ }\href {https://doi.org/10.1038/s41586-023-06910-y} {\bibfield
  {journal} {\bibinfo  {journal} {Nature}\ }\textbf {\bibinfo {volume} {625}},\
  \bibinfo {pages} {673} (\bibinfo {year} {2024})}\BibitemShut {NoStop}%
\bibitem [{\citenamefont {M\"oller}\ \emph {et~al.}(1989)\citenamefont
  {M\"oller}, \citenamefont {Tr{\"{a}}bert}, \citenamefont {Lodwig},
  \citenamefont {Wagner}, \citenamefont {Heckmann}, \citenamefont {Blanke},
  \citenamefont {Livingston},\ and\ \citenamefont {Mokler}}]{Moeller1989}%
  \BibitemOpen
  \bibfield  {author} {\bibinfo {author} {\bibfnamefont {G.}~\bibnamefont
  {M\"oller}}, \bibinfo {author} {\bibfnamefont {E.}~\bibnamefont
  {Tr{\"{a}}bert}}, \bibinfo {author} {\bibfnamefont {V.}~\bibnamefont
  {Lodwig}}, \bibinfo {author} {\bibfnamefont {C.}~\bibnamefont {Wagner}},
  \bibinfo {author} {\bibfnamefont {P.~H.}\ \bibnamefont {Heckmann}}, \bibinfo
  {author} {\bibfnamefont {J.~H.}\ \bibnamefont {Blanke}}, \bibinfo {author}
  {\bibfnamefont {A.~E.}\ \bibnamefont {Livingston}},\ and\ \bibinfo {author}
  {\bibfnamefont {P.~H.}\ \bibnamefont {Mokler}},\ }\bibfield  {title}
  {\bibinfo {title} {Experimental transition probability for the {E1}
  intercombination transition in {B}e-like {Xe$^ {50+}$}},\ }\href
  {https://doi.org/10.1007/BF01438508} {\bibfield  {journal} {\bibinfo
  {journal} {Z. Phys. D}\ }\textbf {\bibinfo {volume} {11}},\ \bibinfo {pages}
  {333} (\bibinfo {year} {1989})}\BibitemShut {NoStop}%
\bibitem [{\citenamefont {B\"uttner}\ \emph {et~al.}(1992)\citenamefont
  {B\"uttner}, \citenamefont {Kraus}, \citenamefont {Schartner}, \citenamefont
  {Folkmann}, \citenamefont {Mokler},\ and\ \citenamefont
  {M\"oller}}]{Buettner1992}%
  \BibitemOpen
  \bibfield  {author} {\bibinfo {author} {\bibfnamefont {R.}~\bibnamefont
  {B\"uttner}}, \bibinfo {author} {\bibfnamefont {B.}~\bibnamefont {Kraus}},
  \bibinfo {author} {\bibfnamefont {K.-H.}\ \bibnamefont {Schartner}}, \bibinfo
  {author} {\bibfnamefont {F.}~\bibnamefont {Folkmann}}, \bibinfo {author}
  {\bibfnamefont {P.~H.}\ \bibnamefont {Mokler}},\ and\ \bibinfo {author}
  {\bibfnamefont {G.}~\bibnamefont {M\"oller}},\ }\bibfield  {title} {\bibinfo
  {title} {{EUV} spectroscopy of beam-foil excited 14.25 {MeV}/u
  {Xe$^{52+}\ldots$Xe$^{49+}$} ions},\ }\href
  {https://doi.org/10.1007/BF01437250} {\bibfield  {journal} {\bibinfo
  {journal} {Z. Phys. D}\ }\textbf {\bibinfo {volume} {22}},\ \bibinfo {pages}
  {693} (\bibinfo {year} {1992})}\BibitemShut {NoStop}%
\bibitem [{\citenamefont {Feili}\ \emph {et~al.}(2005)\citenamefont {Feili},
  \citenamefont {Zimmermann}, \citenamefont {Neacsu}, \citenamefont
  {Bosselmann}, \citenamefont {Schartner}, \citenamefont {Folkmann},
  \citenamefont {Livingston}, \citenamefont {Tr\"{a}bert},\ and\ \citenamefont
  {Mokler}}]{Feili2005}%
  \BibitemOpen
  \bibfield  {author} {\bibinfo {author} {\bibfnamefont {D.}~\bibnamefont
  {Feili}}, \bibinfo {author} {\bibfnamefont {B.}~\bibnamefont {Zimmermann}},
  \bibinfo {author} {\bibfnamefont {C.}~\bibnamefont {Neacsu}}, \bibinfo
  {author} {\bibfnamefont {P.}~\bibnamefont {Bosselmann}}, \bibinfo {author}
  {\bibfnamefont {K.-H.}\ \bibnamefont {Schartner}}, \bibinfo {author}
  {\bibfnamefont {F.}~\bibnamefont {Folkmann}}, \bibinfo {author}
  {\bibfnamefont {A.~E.}\ \bibnamefont {Livingston}}, \bibinfo {author}
  {\bibfnamefont {E.}~\bibnamefont {Tr\"{a}bert}},\ and\ \bibinfo {author}
  {\bibfnamefont {P.~H.}\ \bibnamefont {Mokler}},\ }\bibfield  {title}
  {\bibinfo {title} {{$2s^2\;^1S_0\;\to\;2s2p\;^3P_1$} intercombination
  transition wavelengths in {B}e-like {Ag$^{43+}$}, {Sn$^{46+}$}, and
  {Xe$^{50+}$} ions},\ }\href {https://doi.org/10.1088/0031-8949/71/1/008}
  {\bibfield  {journal} {\bibinfo  {journal} {Phys. Scr.}\ }\textbf {\bibinfo
  {volume} {71}},\ \bibinfo {pages} {48} (\bibinfo {year} {2005})}\BibitemShut
  {NoStop}%
\bibitem [{\citenamefont {Tr\"abert}\ \emph {et~al.}(2003)\citenamefont
  {Tr\"abert}, \citenamefont {Beiersdorfer}, \citenamefont {Lepson},\ and\
  \citenamefont {Chen}}]{Traebert2003}%
  \BibitemOpen
  \bibfield  {author} {\bibinfo {author} {\bibfnamefont {E.}~\bibnamefont
  {Tr\"abert}}, \bibinfo {author} {\bibfnamefont {P.}~\bibnamefont
  {Beiersdorfer}}, \bibinfo {author} {\bibfnamefont {J.~K.}\ \bibnamefont
  {Lepson}},\ and\ \bibinfo {author} {\bibfnamefont {H.}~\bibnamefont {Chen}},\
  }\bibfield  {title} {\bibinfo {title} {Extreme ultraviolet spectra of highly
  charged {X}e ions},\ }\href {https://doi.org/10.1103/PhysRevA.68.042501}
  {\bibfield  {journal} {\bibinfo  {journal} {Phys. Rev. A}\ }\textbf {\bibinfo
  {volume} {68}},\ \bibinfo {pages} {042501} (\bibinfo {year}
  {2003})}\BibitemShut {NoStop}%
\bibitem [{\citenamefont {Bernhardt}\ \emph {et~al.}(2015)\citenamefont
  {Bernhardt}, \citenamefont {Brandau}, \citenamefont {Harman}, \citenamefont
  {Kozhuharov}, \citenamefont {B\"{o}hm}, \citenamefont {Bosch}, \citenamefont
  {Fritzsche}, \citenamefont {Jacobi}, \citenamefont {Kieslich}, \citenamefont
  {Knopp}, \citenamefont {Nolden}, \citenamefont {Shi}, \citenamefont
  {Stachura}, \citenamefont {Steck}, \citenamefont {St\"{o}hlker},
  \citenamefont {Schippers},\ and\ \citenamefont
  {M\"{u}ller}}]{Bernhardt2015a}%
  \BibitemOpen
  \bibfield  {author} {\bibinfo {author} {\bibfnamefont {D.}~\bibnamefont
  {Bernhardt}}, \bibinfo {author} {\bibfnamefont {C.}~\bibnamefont {Brandau}},
  \bibinfo {author} {\bibfnamefont {Z.}~\bibnamefont {Harman}}, \bibinfo
  {author} {\bibfnamefont {C.}~\bibnamefont {Kozhuharov}}, \bibinfo {author}
  {\bibfnamefont {S.}~\bibnamefont {B\"{o}hm}}, \bibinfo {author}
  {\bibfnamefont {F.}~\bibnamefont {Bosch}}, \bibinfo {author} {\bibfnamefont
  {S.}~\bibnamefont {Fritzsche}}, \bibinfo {author} {\bibfnamefont
  {J.}~\bibnamefont {Jacobi}}, \bibinfo {author} {\bibfnamefont
  {S.}~\bibnamefont {Kieslich}}, \bibinfo {author} {\bibfnamefont
  {H.}~\bibnamefont {Knopp}}, \bibinfo {author} {\bibfnamefont
  {F.}~\bibnamefont {Nolden}}, \bibinfo {author} {\bibfnamefont
  {W.}~\bibnamefont {Shi}}, \bibinfo {author} {\bibfnamefont {Z.}~\bibnamefont
  {Stachura}}, \bibinfo {author} {\bibfnamefont {M.}~\bibnamefont {Steck}},
  \bibinfo {author} {\bibfnamefont {T.}~\bibnamefont {St\"{o}hlker}}, \bibinfo
  {author} {\bibfnamefont {S.}~\bibnamefont {Schippers}},\ and\ \bibinfo
  {author} {\bibfnamefont {A.}~\bibnamefont {M\"{u}ller}},\ }\bibfield  {title}
  {\bibinfo {title} {Spectroscopy of berylliumlike xenon ions using
  dielectronic recombination},\ }\href
  {https://doi.org/10.1088/0953-4075/48/14/144008} {\bibfield  {journal}
  {\bibinfo  {journal} {J. Phys. B}\ }\textbf {\bibinfo {volume} {48}},\
  \bibinfo {pages} {144008} (\bibinfo {year} {2015})}\BibitemShut {NoStop}%
\bibitem [{\citenamefont {Cheng}\ \emph {et~al.}(1979)\citenamefont {Cheng},
  \citenamefont {Kim},\ and\ \citenamefont {Desclaux}}]{Cheng1979}%
  \BibitemOpen
  \bibfield  {author} {\bibinfo {author} {\bibfnamefont {K.~T.}\ \bibnamefont
  {Cheng}}, \bibinfo {author} {\bibfnamefont {Y.-K.}\ \bibnamefont {Kim}},\
  and\ \bibinfo {author} {\bibfnamefont {J.~P.}\ \bibnamefont {Desclaux}},\
  }\bibfield  {title} {\bibinfo {title} {Electric dipole, quadrupole and
  magnetic dipole transition probabilities of ions isoelectric to the first-row
  atoms {L}i through {F}},\ }\href
  {https://doi.org/10.1016/0092-640X(79)90006-8} {\bibfield  {journal}
  {\bibinfo  {journal} {At. Data Nucl. Data Tables}\ }\textbf {\bibinfo
  {volume} {24}},\ \bibinfo {pages} {111} (\bibinfo {year} {1979})}\BibitemShut
  {NoStop}%
\bibitem [{\citenamefont {Safronova}\ \emph {et~al.}(1996)\citenamefont
  {Safronova}, \citenamefont {Johnson},\ and\ \citenamefont
  {Safronova}}]{Safronova1996b}%
  \BibitemOpen
  \bibfield  {author} {\bibinfo {author} {\bibfnamefont {M.~S.}\ \bibnamefont
  {Safronova}}, \bibinfo {author} {\bibfnamefont {W.~R.}\ \bibnamefont
  {Johnson}},\ and\ \bibinfo {author} {\bibfnamefont {U.~I.}\ \bibnamefont
  {Safronova}},\ }\bibfield  {title} {\bibinfo {title} {Relativistic many-body
  calculations of the energies of n=2 states for the berylliumlike
  isoelectronic sequence},\ }\href {https://doi.org/10.1103/PhysRevA.53.4036}
  {\bibfield  {journal} {\bibinfo  {journal} {Phys. Rev. A}\ }\textbf {\bibinfo
  {volume} {53}},\ \bibinfo {pages} {4036} (\bibinfo {year}
  {1996})}\BibitemShut {NoStop}%
\bibitem [{\citenamefont {Gu}(2005)}]{Gu2005a}%
  \BibitemOpen
  \bibfield  {author} {\bibinfo {author} {\bibfnamefont {M.~F.}\ \bibnamefont
  {Gu}},\ }\bibfield  {title} {\bibinfo {title} {Energies of {$1s^22l^q (1\leq
  q\leq8)$} states for {$Z\leq60$} with a combined configuration interaction
  and many-body perturbation theory approach},\ }\href
  {https://doi.org/https://doi.org/10.1016/j.adt.2005.02.004} {\bibfield
  {journal} {\bibinfo  {journal} {At. Data Nucl. Data Tables}\ }\textbf
  {\bibinfo {volume} {89}},\ \bibinfo {pages} {267} (\bibinfo {year}
  {2005})}\BibitemShut {NoStop}%
\bibitem [{\citenamefont {Cheng}\ \emph {et~al.}(2008)\citenamefont {Cheng},
  \citenamefont {Chen},\ and\ \citenamefont {Johnson}}]{Cheng2008a}%
  \BibitemOpen
  \bibfield  {author} {\bibinfo {author} {\bibfnamefont {K.~T.}\ \bibnamefont
  {Cheng}}, \bibinfo {author} {\bibfnamefont {M.~H.}\ \bibnamefont {Chen}},\
  and\ \bibinfo {author} {\bibfnamefont {W.~R.}\ \bibnamefont {Johnson}},\
  }\bibfield  {title} {\bibinfo {title} {Hyperfine quenching of the
  $2s\,2p\;{^3}{P}_0$ state of berylliumlike ions},\ }\href
  {https://doi.org/10.1103/PhysRevA.77.052504} {\bibfield  {journal} {\bibinfo
  {journal} {Phys. Rev. A}\ }\textbf {\bibinfo {volume} {77}},\ \bibinfo
  {pages} {052504} (\bibinfo {year} {2008})}\BibitemShut {NoStop}%
\bibitem [{\citenamefont {Malyshev}\ \emph {et~al.}(2014)\citenamefont
  {Malyshev}, \citenamefont {Volotka}, \citenamefont {Glazov}, \citenamefont
  {Tupitsyn}, \citenamefont {Shabaev},\ and\ \citenamefont
  {Plunien}}]{Malyshev2014}%
  \BibitemOpen
  \bibfield  {author} {\bibinfo {author} {\bibfnamefont {A.~V.}\ \bibnamefont
  {Malyshev}}, \bibinfo {author} {\bibfnamefont {A.~V.}\ \bibnamefont
  {Volotka}}, \bibinfo {author} {\bibfnamefont {D.~A.}\ \bibnamefont {Glazov}},
  \bibinfo {author} {\bibfnamefont {I.~I.}\ \bibnamefont {Tupitsyn}}, \bibinfo
  {author} {\bibfnamefont {V.~M.}\ \bibnamefont {Shabaev}},\ and\ \bibinfo
  {author} {\bibfnamefont {G.}~\bibnamefont {Plunien}},\ }\bibfield  {title}
  {\bibinfo {title} {{QED} calculation of the ground-state energy of
  berylliumlike ions},\ }\href {https://doi.org/10.1103/PhysRevA.90.062517}
  {\bibfield  {journal} {\bibinfo  {journal} {Phys. Rev. A}\ }\textbf {\bibinfo
  {volume} {90}},\ \bibinfo {pages} {062517} (\bibinfo {year}
  {2014})}\BibitemShut {NoStop}%
\bibitem [{\citenamefont {Malyshev}\ \emph {et~al.}(2015)\citenamefont
  {Malyshev}, \citenamefont {Volotka}, \citenamefont {Glazov}, \citenamefont
  {Tupitsyn}, \citenamefont {Shabaev},\ and\ \citenamefont
  {Plunien}}]{Malyshev2015}%
  \BibitemOpen
  \bibfield  {author} {\bibinfo {author} {\bibfnamefont {A.~V.}\ \bibnamefont
  {Malyshev}}, \bibinfo {author} {\bibfnamefont {A.~V.}\ \bibnamefont
  {Volotka}}, \bibinfo {author} {\bibfnamefont {D.~A.}\ \bibnamefont {Glazov}},
  \bibinfo {author} {\bibfnamefont {I.~I.}\ \bibnamefont {Tupitsyn}}, \bibinfo
  {author} {\bibfnamefont {V.~M.}\ \bibnamefont {Shabaev}},\ and\ \bibinfo
  {author} {\bibfnamefont {G.}~\bibnamefont {Plunien}},\ }\bibfield  {title}
  {\bibinfo {title} {Ionization energies along beryllium isoelectronic
  sequence},\ }\href {https://doi.org/10.1103/PhysRevA.92.012514} {\bibfield
  {journal} {\bibinfo  {journal} {Phys. Rev. A}\ }\textbf {\bibinfo {volume}
  {92}},\ \bibinfo {pages} {012514} (\bibinfo {year} {2015})}\BibitemShut
  {NoStop}%
\bibitem [{\citenamefont {Kaygorodov}\ \emph {et~al.}(2019)\citenamefont
  {Kaygorodov}, \citenamefont {Kozhedub}, \citenamefont {Tupitsyn},
  \citenamefont {Malyshev}, \citenamefont {Glazov}, \citenamefont {Plunien},\
  and\ \citenamefont {Shabaev}}]{Kaygorodov2019}%
  \BibitemOpen
  \bibfield  {author} {\bibinfo {author} {\bibfnamefont {M.~Y.}\ \bibnamefont
  {Kaygorodov}}, \bibinfo {author} {\bibfnamefont {Y.~S.}\ \bibnamefont
  {Kozhedub}}, \bibinfo {author} {\bibfnamefont {I.~I.}\ \bibnamefont
  {Tupitsyn}}, \bibinfo {author} {\bibfnamefont {A.~V.}\ \bibnamefont
  {Malyshev}}, \bibinfo {author} {\bibfnamefont {D.~A.}\ \bibnamefont
  {Glazov}}, \bibinfo {author} {\bibfnamefont {G.}~\bibnamefont {Plunien}},\
  and\ \bibinfo {author} {\bibfnamefont {V.~M.}\ \bibnamefont {Shabaev}},\
  }\bibfield  {title} {\bibinfo {title} {Relativistic calculations of the
  ground and inner-{$L$}-shell excited energy levels of berylliumlike ions},\
  }\href {https://doi.org/10.1103/PhysRevA.99.032505} {\bibfield  {journal}
  {\bibinfo  {journal} {Phys. Rev. A}\ }\textbf {\bibinfo {volume} {99}},\
  \bibinfo {pages} {032505} (\bibinfo {year} {2019})}\BibitemShut {NoStop}%
\bibitem [{\citenamefont {Malyshev}\ \emph {et~al.}(2023)\citenamefont
  {Malyshev}, \citenamefont {Kozhedub},\ and\ \citenamefont
  {Shabaev}}]{Malyshev2023}%
  \BibitemOpen
  \bibfield  {author} {\bibinfo {author} {\bibfnamefont {A.~V.}\ \bibnamefont
  {Malyshev}}, \bibinfo {author} {\bibfnamefont {Y.~S.}\ \bibnamefont
  {Kozhedub}},\ and\ \bibinfo {author} {\bibfnamefont {V.~M.}\ \bibnamefont
  {Shabaev}},\ }\bibfield  {title} {\bibinfo {title} {Ab initio calculations of
  the {$2p_{3/2} \to 2s$} transition in \mbox{He-,} {Li-,} and {Be-}like
  uranium},\ }\href {https://doi.org/10.1103/PhysRevA.107.042806} {\bibfield
  {journal} {\bibinfo  {journal} {Phys. Rev. A}\ }\textbf {\bibinfo {volume}
  {107}},\ \bibinfo {pages} {042806} (\bibinfo {year} {2023})}\BibitemShut
  {NoStop}%
\bibitem [{\citenamefont {Malyshev}\ \emph {et~al.}(2024)\citenamefont
  {Malyshev}, \citenamefont {Kozhedub}, \citenamefont {Shabaev},\ and\
  \citenamefont {Tupitsyn}}]{Malyshev2024}%
  \BibitemOpen
  \bibfield  {author} {\bibinfo {author} {\bibfnamefont {A.~V.}\ \bibnamefont
  {Malyshev}}, \bibinfo {author} {\bibfnamefont {Y.~S.}\ \bibnamefont
  {Kozhedub}}, \bibinfo {author} {\bibfnamefont {V.~M.}\ \bibnamefont
  {Shabaev}},\ and\ \bibinfo {author} {\bibfnamefont {I.~I.}\ \bibnamefont
  {Tupitsyn}},\ }\bibfield  {title} {\bibinfo {title} {{QED} calculations of
  intra-{L}-shell singly excited states in {B}e-like ions},\ }\href
  {https://doi.org/10.1103/PhysRevA.110.062824} {\bibfield  {journal} {\bibinfo
   {journal} {Phys. Rev. A}\ }\textbf {\bibinfo {volume} {110}},\ \bibinfo
  {pages} {062824} (\bibinfo {year} {2024})}\BibitemShut {NoStop}%
\bibitem [{\citenamefont {Krantz}\ \emph {et~al.}(2021)\citenamefont {Krantz},
  \citenamefont {Andelkovic}, \citenamefont {Brandau}, \citenamefont
  {Dimopoulou}, \citenamefont {Geithner}, \citenamefont {Hackler},
  \citenamefont {Hannen}, \citenamefont {Herfurth}, \citenamefont {Hess},
  \citenamefont {Lestinsky}, \citenamefont {Menz}, \citenamefont {Reiter},
  \citenamefont {Ro{\ss}bach}, \citenamefont {Schippers}, \citenamefont
  {Schroeder}, \citenamefont {T{\"{a}}schner}, \citenamefont {Vorobjev},
  \citenamefont {Weinheimer},\ and\ \citenamefont {Winzen}}]{Krantz2021}%
  \BibitemOpen
  \bibfield  {author} {\bibinfo {author} {\bibfnamefont {C.}~\bibnamefont
  {Krantz}}, \bibinfo {author} {\bibfnamefont {Z.}~\bibnamefont {Andelkovic}},
  \bibinfo {author} {\bibfnamefont {C.}~\bibnamefont {Brandau}}, \bibinfo
  {author} {\bibfnamefont {C.}~\bibnamefont {Dimopoulou}}, \bibinfo {author}
  {\bibfnamefont {W.}~\bibnamefont {Geithner}}, \bibinfo {author}
  {\bibfnamefont {T.}~\bibnamefont {Hackler}}, \bibinfo {author} {\bibfnamefont
  {V.}~\bibnamefont {Hannen}}, \bibinfo {author} {\bibfnamefont
  {F.}~\bibnamefont {Herfurth}}, \bibinfo {author} {\bibfnamefont
  {R.}~\bibnamefont {Hess}}, \bibinfo {author} {\bibfnamefont {M.}~\bibnamefont
  {Lestinsky}}, \bibinfo {author} {\bibfnamefont {E.}~\bibnamefont {Menz}},
  \bibinfo {author} {\bibfnamefont {A.}~\bibnamefont {Reiter}}, \bibinfo
  {author} {\bibfnamefont {J.}~\bibnamefont {Ro{\ss}bach}}, \bibinfo {author}
  {\bibfnamefont {S.}~\bibnamefont {Schippers}}, \bibinfo {author}
  {\bibfnamefont {C.}~\bibnamefont {Schroeder}}, \bibinfo {author}
  {\bibfnamefont {A.}~\bibnamefont {T{\"{a}}schner}}, \bibinfo {author}
  {\bibfnamefont {G.}~\bibnamefont {Vorobjev}}, \bibinfo {author}
  {\bibfnamefont {C.}~\bibnamefont {Weinheimer}},\ and\ \bibinfo {author}
  {\bibfnamefont {D.}~\bibnamefont {Winzen}},\ }\bibfield  {title} {\bibinfo
  {title} {Recommissioning of the {CRYRING@ESR} electron cooler},\ }in\ \href
  {https://doi.org/10.18429/JACoW-IPAC2021-TUPAB178} {\emph {\bibinfo
  {booktitle} {Proc. {IPAC'21}}}},\ \bibinfo {series and number} {\bibinfo
  {series} {International Particle Accelerator Conference}\ No.~\bibinfo
  {number} {12}}\ (\bibinfo  {publisher} {JACoW Publishing, Geneva,
  Switzerland},\ \bibinfo {year} {2021})\ pp.\ \bibinfo {pages}
  {1816--1818}\BibitemShut {NoStop}%
\bibitem [{\citenamefont {Wang}\ \emph {et~al.}(2024)\citenamefont {Wang},
  \citenamefont {Brandau}, \citenamefont {Harman}, \citenamefont {Fritzsche},
  \citenamefont {Fuchs}, \citenamefont {Kozhuharov}, \citenamefont {M\"uller},
  \citenamefont {Steck},\ and\ \citenamefont {Schippers}}]{Wang2024}%
  \BibitemOpen
  \bibfield  {author} {\bibinfo {author} {\bibfnamefont {S.~X.}\ \bibnamefont
  {Wang}}, \bibinfo {author} {\bibfnamefont {C.}~\bibnamefont {Brandau}},
  \bibinfo {author} {\bibfnamefont {Z.}~\bibnamefont {Harman}}, \bibinfo
  {author} {\bibfnamefont {S.}~\bibnamefont {Fritzsche}}, \bibinfo {author}
  {\bibfnamefont {S.}~\bibnamefont {Fuchs}}, \bibinfo {author} {\bibfnamefont
  {C.}~\bibnamefont {Kozhuharov}}, \bibinfo {author} {\bibfnamefont
  {A.}~\bibnamefont {M\"uller}}, \bibinfo {author} {\bibfnamefont
  {M.}~\bibnamefont {Steck}},\ and\ \bibinfo {author} {\bibfnamefont
  {S.}~\bibnamefont {Schippers}},\ }\bibfield  {title} {\bibinfo {title} {Breit
  interaction in dielectronic recombination of hydrogenlike xenon ions:
  storage-ring experiment and theory},\ }\href
  {https://doi.org/10.1140/epjd/s10053-024-00914-7} {\bibfield  {journal}
  {\bibinfo  {journal} {Eur. Phys. J. D}\ }\textbf {\bibinfo {volume} {78}},\
  \bibinfo {pages} {122} (\bibinfo {year} {2024})}\BibitemShut {NoStop}%
\bibitem [{\citenamefont {Th{\"{u}}mmler}\ \emph {et~al.}(2009)\citenamefont
  {Th{\"{u}}mmler}, \citenamefont {Marx},\ and\ \citenamefont
  {Weinheimer}}]{Thuemmler2009}%
  \BibitemOpen
  \bibfield  {author} {\bibinfo {author} {\bibfnamefont {T.}~\bibnamefont
  {Th{\"{u}}mmler}}, \bibinfo {author} {\bibfnamefont {R.}~\bibnamefont
  {Marx}},\ and\ \bibinfo {author} {\bibfnamefont {C.}~\bibnamefont
  {Weinheimer}},\ }\bibfield  {title} {\bibinfo {title} {Precision high voltage
  divider for the {KATRIN} experiment},\ }\href
  {https://doi.org/10.1088/1367-2630/11/10/103007} {\bibfield  {journal}
  {\bibinfo  {journal} {New J. Phys.}\ }\textbf {\bibinfo {volume} {11}},\
  \bibinfo {pages} {103007} (\bibinfo {year} {2009})}\BibitemShut {NoStop}%
\bibitem [{\citenamefont {Rest}\ \emph {et~al.}(2020)\citenamefont {Rest},
  \citenamefont {Winzen}, \citenamefont {Hannen},\ and\ \citenamefont
  {Weinheimer}}]{Rest2020}%
  \BibitemOpen
  \bibfield  {author} {\bibinfo {author} {\bibfnamefont {O.}~\bibnamefont
  {Rest}}, \bibinfo {author} {\bibfnamefont {D.}~\bibnamefont {Winzen}},
  \bibinfo {author} {\bibfnamefont {V.}~\bibnamefont {Hannen}},\ and\ \bibinfo
  {author} {\bibfnamefont {C.}~\bibnamefont {Weinheimer}},\ }\bibfield  {title}
  {\bibinfo {title} {Absolute calibration of a ppm-precise {HV} divider for the
  electron cooler of the ion storage ring {CRYRING@ESR}},\ }in\ \href
  {https://doi.org/10.1007/978-3-030-31680-8_142} {\emph {\bibinfo {booktitle}
  {Proceedings of the 21st International Symposium on High Voltage
  Engineering}}},\ \bibinfo {editor} {edited by\ \bibinfo {editor}
  {\bibfnamefont {B.}~\bibnamefont {Németh}}}\ (\bibinfo  {publisher}
  {Springer International Publishing},\ \bibinfo {address} {Cham},\ \bibinfo
  {year} {2020})\ pp.\ \bibinfo {pages} {1500--1512}\BibitemShut {NoStop}%
\bibitem [{\citenamefont {B\"{o}hm}\ \emph {et~al.}(2001)\citenamefont
  {B\"{o}hm}, \citenamefont {Schippers}, \citenamefont {Shi}, \citenamefont
  {M\"{u}ller}, \citenamefont {Djuri\'{c}}, \citenamefont {Dunn}, \citenamefont
  {Zong}, \citenamefont {Jelenkovi\'{c}}, \citenamefont {Danared},
  \citenamefont {Ekl\"{o}w}, \citenamefont {Glans},\ and\ \citenamefont
  {Schuch}}]{Boehm2001b}%
  \BibitemOpen
  \bibfield  {author} {\bibinfo {author} {\bibfnamefont {S.}~\bibnamefont
  {B\"{o}hm}}, \bibinfo {author} {\bibfnamefont {S.}~\bibnamefont {Schippers}},
  \bibinfo {author} {\bibfnamefont {W.}~\bibnamefont {Shi}}, \bibinfo {author}
  {\bibfnamefont {A.}~\bibnamefont {M\"{u}ller}}, \bibinfo {author}
  {\bibfnamefont {N.}~\bibnamefont {Djuri\'{c}}}, \bibinfo {author}
  {\bibfnamefont {G.~H.}\ \bibnamefont {Dunn}}, \bibinfo {author}
  {\bibfnamefont {W.}~\bibnamefont {Zong}}, \bibinfo {author} {\bibfnamefont
  {B.}~\bibnamefont {Jelenkovi\'{c}}}, \bibinfo {author} {\bibfnamefont
  {H.}~\bibnamefont {Danared}}, \bibinfo {author} {\bibfnamefont
  {N.}~\bibnamefont {Ekl\"{o}w}}, \bibinfo {author} {\bibfnamefont
  {P.}~\bibnamefont {Glans}},\ and\ \bibinfo {author} {\bibfnamefont
  {R.}~\bibnamefont {Schuch}},\ }\bibfield  {title} {\bibinfo {title}
  {Influence of electromagnetic fields on the dielectronic recombination of
  {N}e$^{7+}$ ions},\ }\href {https://doi.org/10.1103/PhysRevA.64.032707}
  {\bibfield  {journal} {\bibinfo  {journal} {Phys. Rev. A}\ }\textbf {\bibinfo
  {volume} {64}},\ \bibinfo {pages} {032707} (\bibinfo {year}
  {2001})}\BibitemShut {NoStop}%
\bibitem [{\citenamefont {Badnell}(2011)}]{Badnell2011b}%
  \BibitemOpen
  \bibfield  {author} {\bibinfo {author} {\bibfnamefont {N.~R.}\ \bibnamefont
  {Badnell}},\ }\bibfield  {title} {\bibinfo {title} {A {B}reit-{P}auli
  distorted wave implementation for \textsc{autostructure}},\ }\href
  {https://doi.org/10.1016/j.cpc.2011.03.023} {\bibfield  {journal} {\bibinfo
  {journal} {Comput. Phys. Commun.}\ }\textbf {\bibinfo {volume} {182}},\
  \bibinfo {pages} {1528} (\bibinfo {year} {2011})}\BibitemShut {NoStop}%
\bibitem [{\citenamefont {Iafrate}\ and\ \citenamefont
  {Mendelsohn}(1969)}]{Iafrate1969}%
  \BibitemOpen
  \bibfield  {author} {\bibinfo {author} {\bibfnamefont {G.~J.}\ \bibnamefont
  {Iafrate}}\ and\ \bibinfo {author} {\bibfnamefont {L.~B.}\ \bibnamefont
  {Mendelsohn}},\ }\bibfield  {title} {\bibinfo {title} {High-order
  perturbation theory for the bound states of an electron in a screened
  {C}oulomb potential},\ }\href {https://doi.org/10.1103/PhysRev.182.244}
  {\bibfield  {journal} {\bibinfo  {journal} {Phys. Rev.}\ }\textbf {\bibinfo
  {volume} {182}},\ \bibinfo {pages} {244} (\bibinfo {year}
  {1969})}\BibitemShut {NoStop}%
\bibitem [{\citenamefont {Badnell}\ \emph {et~al.}(2004)\citenamefont
  {Badnell}, \citenamefont {Mitnik}, \citenamefont {Pindzola}, \citenamefont
  {Loch},\ and\ \citenamefont {Abdel-Naby}}]{Badnell2004a}%
  \BibitemOpen
  \bibfield  {author} {\bibinfo {author} {\bibfnamefont {N.~R.}\ \bibnamefont
  {Badnell}}, \bibinfo {author} {\bibfnamefont {D.~M.}\ \bibnamefont {Mitnik}},
  \bibinfo {author} {\bibfnamefont {M.~S.}\ \bibnamefont {Pindzola}}, \bibinfo
  {author} {\bibfnamefont {S.~D.}\ \bibnamefont {Loch}},\ and\ \bibinfo
  {author} {\bibfnamefont {S.~A.}\ \bibnamefont {Abdel-Naby}},\ }\bibfield
  {title} {\bibinfo {title} {Dielectronic recombination of {P}b$^{79+}$ via
  high angular momenta},\ }\href {https://doi.org/10.1103/PhysRevA.70.054701}
  {\bibfield  {journal} {\bibinfo  {journal} {Phys. Rev. A}\ }\textbf {\bibinfo
  {volume} {70}},\ \bibinfo {pages} {054701} (\bibinfo {year}
  {2004})}\BibitemShut {NoStop}%
\bibitem [{\citenamefont {Shabaeva}(2003)}]{Shabaeva2003}%
  \BibitemOpen
  \bibfield  {author} {\bibinfo {author} {\bibfnamefont {M.~B.}\ \bibnamefont
  {Shabaeva}},\ }\bibfield  {title} {\bibinfo {title} {Calculation of the
  binding energy of a highly excited electron in heavy {B}e-like ions},\ }\href
  {https://doi.org/10.1134/1.1595203} {\bibfield  {journal} {\bibinfo
  {journal} {Optics and Spectroscopy}\ }\textbf {\bibinfo {volume} {95}},\
  \bibinfo {pages} {1} (\bibinfo {year} {2003})}\BibitemShut {NoStop}%
\bibitem [{\citenamefont {Brandau}\ \emph {et~al.}(2002)\citenamefont
  {Brandau}, \citenamefont {Bartsch}, \citenamefont {Hoffknecht}, \citenamefont
  {Knopp}, \citenamefont {Schippers}, \citenamefont {Shi}, \citenamefont
  {M\"{u}ller}, \citenamefont {Gr\"{u}n}, \citenamefont {Scheid}, \citenamefont
  {Steih}, \citenamefont {Bosch}, \citenamefont {Franzke}, \citenamefont
  {Kozhuharov}, \citenamefont {Mokler}, \citenamefont {Nolden}, \citenamefont
  {Steck}, \citenamefont {St\"{o}hlker},\ and\ \citenamefont
  {Stachura}}]{Brandau2002a}%
  \BibitemOpen
  \bibfield  {author} {\bibinfo {author} {\bibfnamefont {C.}~\bibnamefont
  {Brandau}}, \bibinfo {author} {\bibfnamefont {T.}~\bibnamefont {Bartsch}},
  \bibinfo {author} {\bibfnamefont {A.}~\bibnamefont {Hoffknecht}}, \bibinfo
  {author} {\bibfnamefont {H.}~\bibnamefont {Knopp}}, \bibinfo {author}
  {\bibfnamefont {S.}~\bibnamefont {Schippers}}, \bibinfo {author}
  {\bibfnamefont {W.}~\bibnamefont {Shi}}, \bibinfo {author} {\bibfnamefont
  {A.}~\bibnamefont {M\"{u}ller}}, \bibinfo {author} {\bibfnamefont
  {N.}~\bibnamefont {Gr\"{u}n}}, \bibinfo {author} {\bibfnamefont
  {W.}~\bibnamefont {Scheid}}, \bibinfo {author} {\bibfnamefont
  {T.}~\bibnamefont {Steih}}, \bibinfo {author} {\bibfnamefont
  {F.}~\bibnamefont {Bosch}}, \bibinfo {author} {\bibfnamefont
  {B.}~\bibnamefont {Franzke}}, \bibinfo {author} {\bibfnamefont
  {C.}~\bibnamefont {Kozhuharov}}, \bibinfo {author} {\bibfnamefont {P.~H.}\
  \bibnamefont {Mokler}}, \bibinfo {author} {\bibfnamefont {F.}~\bibnamefont
  {Nolden}}, \bibinfo {author} {\bibfnamefont {M.}~\bibnamefont {Steck}},
  \bibinfo {author} {\bibfnamefont {T.}~\bibnamefont {St\"{o}hlker}},\ and\
  \bibinfo {author} {\bibfnamefont {Z.}~\bibnamefont {Stachura}},\ }\bibfield
  {title} {\bibinfo {title} {High {R}ydberg resonances in dielectronic
  recombination of {Pb$^{79+}$}},\ }\href
  {https://doi.org/10.1103/PhysRevLett.89.053201} {\bibfield  {journal}
  {\bibinfo  {journal} {Phys. Rev. Lett.}\ }\textbf {\bibinfo {volume} {89}},\
  \bibinfo {pages} {053201} (\bibinfo {year} {2002})}\BibitemShut {NoStop}%
\bibitem [{\citenamefont {Shi}\ \emph {et~al.}(2002)\citenamefont {Shi},
  \citenamefont {Bartsch}, \citenamefont {B\"{o}hme}, \citenamefont {Brandau},
  \citenamefont {Hoffknecht}, \citenamefont {Knopp}, \citenamefont {Schippers},
  \citenamefont {M\"{u}ller}, \citenamefont {Kozhuharov}, \citenamefont
  {Beckert}, \citenamefont {Bosch}, \citenamefont {Franzke}, \citenamefont
  {Mokler}, \citenamefont {Nolden}, \citenamefont {Steck}, \citenamefont
  {St\"{o}hlker},\ and\ \citenamefont {Stachura}}]{Shi2002a}%
  \BibitemOpen
  \bibfield  {author} {\bibinfo {author} {\bibfnamefont {W.}~\bibnamefont
  {Shi}}, \bibinfo {author} {\bibfnamefont {T.}~\bibnamefont {Bartsch}},
  \bibinfo {author} {\bibfnamefont {C.}~\bibnamefont {B\"{o}hme}}, \bibinfo
  {author} {\bibfnamefont {C.}~\bibnamefont {Brandau}}, \bibinfo {author}
  {\bibfnamefont {A.}~\bibnamefont {Hoffknecht}}, \bibinfo {author}
  {\bibfnamefont {H.}~\bibnamefont {Knopp}}, \bibinfo {author} {\bibfnamefont
  {S.}~\bibnamefont {Schippers}}, \bibinfo {author} {\bibfnamefont
  {A.}~\bibnamefont {M\"{u}ller}}, \bibinfo {author} {\bibfnamefont
  {C.}~\bibnamefont {Kozhuharov}}, \bibinfo {author} {\bibfnamefont
  {K.}~\bibnamefont {Beckert}}, \bibinfo {author} {\bibfnamefont
  {F.}~\bibnamefont {Bosch}}, \bibinfo {author} {\bibfnamefont
  {B.}~\bibnamefont {Franzke}}, \bibinfo {author} {\bibfnamefont {P.~H.}\
  \bibnamefont {Mokler}}, \bibinfo {author} {\bibfnamefont {F.}~\bibnamefont
  {Nolden}}, \bibinfo {author} {\bibfnamefont {M.}~\bibnamefont {Steck}},
  \bibinfo {author} {\bibfnamefont {T.}~\bibnamefont {St\"{o}hlker}},\ and\
  \bibinfo {author} {\bibfnamefont {Z.}~\bibnamefont {Stachura}},\ }\bibfield
  {title} {\bibinfo {title} {Rate enhancement in the recombination of
  {Bi$^{80+}$} ions with electrons},\ }\href
  {https://doi.org/10.1103/PhysRevA.66.022718} {\bibfield  {journal} {\bibinfo
  {journal} {Phys. Rev. A}\ }\textbf {\bibinfo {volume} {66}},\ \bibinfo
  {pages} {022718} (\bibinfo {year} {2002})}\BibitemShut {NoStop}%
\bibitem [{\citenamefont {Trageser}\ \emph {et~al.}(2015)\citenamefont
  {Trageser}, \citenamefont {Brandau}, \citenamefont {Kozhuharov},
  \citenamefont {Litvinov}, \citenamefont {M\"{u}ller}, \citenamefont {Nolden},
  \citenamefont {Sanjari},\ and\ \citenamefont {St\"{o}hlker}}]{Trageser2015}%
  \BibitemOpen
  \bibfield  {author} {\bibinfo {author} {\bibfnamefont {C.}~\bibnamefont
  {Trageser}}, \bibinfo {author} {\bibfnamefont {C.}~\bibnamefont {Brandau}},
  \bibinfo {author} {\bibfnamefont {C.}~\bibnamefont {Kozhuharov}}, \bibinfo
  {author} {\bibfnamefont {Y.~A.}\ \bibnamefont {Litvinov}}, \bibinfo {author}
  {\bibfnamefont {A.}~\bibnamefont {M\"{u}ller}}, \bibinfo {author}
  {\bibfnamefont {F.}~\bibnamefont {Nolden}}, \bibinfo {author} {\bibfnamefont
  {S.}~\bibnamefont {Sanjari}},\ and\ \bibinfo {author} {\bibfnamefont
  {T.}~\bibnamefont {St\"{o}hlker}},\ }\bibfield  {title} {\bibinfo {title} {A
  new data acquisition system for {S}chottky signals in atomic physics
  experiments at {GSI}'s and {FAIR}'s storage rings},\ }\href
  {https://doi.org/10.1088/0031-8949/2015/T166/014062} {\bibfield  {journal}
  {\bibinfo  {journal} {Phys. Scr.}\ }\textbf {\bibinfo {volume} {T166}},\
  \bibinfo {pages} {014062} (\bibinfo {year} {2015})}\BibitemShut {NoStop}%
\end{thebibliography}

%
\appendix*

\onecolumngrid
\section{End Matter}
\twocolumngrid
\textit{Appendix: Uncertainties due to electron-beam space charge, cooling force, and continuum lowering} --- The space charge of the electron beam influences the electron energy and, therefore, needs to be considered in the discussion of the uncertainty of the experimental energy scale. The cooling energy $E_\mathrm{cool} = -e[U_c-U_\mathrm{sc}(0)]$ is determined by the (negative) electron-cooler cathode voltage $U_c$ at cooling and the negative space-charge potential $U_\mathrm{sc}(U_d)$ on the axis of the electron beam, which (via $\beta_e$) depends on the detuning voltage $U_d$. For $U_d\neq0$~V, the laboratory electron energy is $E_e=E_\mathrm{cool}+E_d = -e[U_c+U_d-U_\mathrm{sc}(U_d)]$ and, thus, $E_d = -e[U_d-\delta U_\mathrm{sc}(U_d)]$ with $\delta U_\mathrm{sc}(U_d)= U_\mathrm{sc}(U_d)-U_\mathrm{sc}(0)$.

The space-charge potential can be computed  \cite{Brandau2025,Boehm2001b} from $\beta_e$, the electron current $I_e=39.3$~mA, and the ratio $r_t/r_e$, where $r_t=5$~cm is the radius of the conducting vacuum tube that surrounds the electron beam with radius $r_e=r_c\sqrt{\zeta} = 1.15$~cm. The latter was calculated from the cooler cathode radius $r_c = 0.2$~cm and the electron-beam expansion factor $\zeta=33$ \cite{Danared1994}. These values result in $U_\mathrm{sc}\approx-30.0$~V and $U_\mathrm{sc}\approx-31.7$~V at $U_d=0$~V and $U_d=+650$~V, respectively, corresponding to $\delta U_\mathrm{sc} \approx -1.7$~V at $E_\mathrm{cm}\approx 17.9$~eV.  

In order to assess the uncertainty of the space charge potential we experimentally determined $U_\mathrm{sc}(0)$ by stepwise reducing the electron density and readjusting $U_c$ such that the velocity of the cooled ion beam stayed constant. By extrapolating these readjustments to zero electron density we arrived at a value of $-27$~V for $U_\mathrm{sc}(0)$, which deviates from our calculated value of $-30$~eV by 10\%. We therefore assign a 10\% uncertainty to $\delta U_\mathrm{sc}$.

The cooling force accelerates or decelerates the ions towards the electron velocity. Thus, in principle, it can change the ion energy during the periods when the electron energy is detuned from the cooling energy. This effect can become considerable for small detunings \cite{Fogle2003a,Shi2002a} when ion beam is uncooled in between two successive measurement steps. In the present work, where $E_\mathrm{cm}$ is comparatively large and where we have used intermittent cooling, the drag effect is very small. Using a similar procedure as in \cite{Fogle2003a,Shi2002a} we estimate that it leads to a change of $E_\mathrm{cool}$ by less than 50~meV, which is well within its uncertainty. This conclusion is also supported by the Schottky ion-beam analysis, where the revolution frequencies of the stored ions are monitored with a pick-up electrode \cite{Trageser2015}. The Schottky spectra did not indicate any deviation of the mean ion-beam velocity from its cooling value during the voltage scanning.

The influence of continuum lowering on the DR resonance positions can be estimated by accounting for the Debye-H\"uckel screening length in a plasma,  
\begin{equation}
D=\sqrt{\frac{\varepsilon_0kT}{n_ee^2}},
\end{equation}
which depends on the electron temperature $kT$ and the electron density $n_e$. $\varepsilon_0$ and $e$ denote the vacuum electric permittivity and the elementary charge, respectively. In our experiment $kT\approx 10^{-3}$~eV and $n_e=10^7$~cm$^{-3}$. These values yield $D= 7.433\cdot10^{-5}$~m.

The hydrogenic atomic level energies in a screened Coulomb potential (Yukawa potential) can be calculated perturbatively \cite{Iafrate1969},  where the expansion parameter is $\epsilon = a_0/D$ and $a_0\approx 5.292\cdot10^{-11}$~m is the Bohr radius. Accordingly, the energy $E_{n,l}$ of a hydrogenic level up to order $\epsilon^2$ is 
\begin{equation}
\frac{E_{n,l}}{E_h} =-\frac{q^2}{2n^2} +q\epsilon -\frac{1}{4}\left[3n^2-l(l+1)\right]\epsilon^2
\end{equation}
with the  Hartree energy  $E_h\approx27.211$~eV.
Inserting $\epsilon = a_0/D \approx 7.116\cdot 10^{-7}$ and $q=78$, the first order correction amounts to $q\epsilon E_h \approx 0.0015$~meV. We conclude that the influence of continuum lowering is negligible compared to our $\pm$30-meV total systematic uncertainty.

\end{document}